\documentclass[pdflatex,twocolumn,sn-nature]{sn-jnl}% Style for submissions to Nature Portfolio journals
\usepackage{graphicx}%
\usepackage{multirow}%
\usepackage{amsmath,amssymb,amsfonts}%
\usepackage{amsthm}%
\usepackage{mathrsfs}%
\usepackage[title]{appendix}%
\usepackage{xcolor}%
\usepackage{textcomp}%
\usepackage{manyfoot}%
\usepackage{booktabs}%
\usepackage{algorithm}%
\usepackage{algorithmicx}%
\usepackage{algpseudocode}%
\usepackage{listings}%
\usepackage{bm}% bold math
\usepackage{tabularx}

\def \k{{\mathbf k}}

\begin{document}

\title[Article Title]{ Magnetization Dependent In-plane Anomalous Hall Effect in a Low-dimensional System }


\author[1]{\fnm{I-Hsuan} \sur{Kao}}
\equalcont{These authors contributed equally to this work.}
\author[1]{\fnm{Ravi Kumar} \sur{Bandapelli}}
\equalcont{These authors contributed equally to this work.}
\author[1]{\fnm{Zhenhong} \sur{Cui}}
\author[1]{\fnm{Shuchen} \sur{Zhang}}
\author[2]{\fnm{Jian} \sur{Tang}}
\author[3]{\fnm{Tiema} \sur{Qian}}
\author[1]{\fnm{Souvik} \sur{Sasmal}}
\author[1]{\fnm{Aalok} \sur{Tiwari}}
\author[1]{\fnm{Mei-Tung} \sur{Chen}}

\author[4]{\fnm{Rahul} \sur{Rao}}

\author[5]{\fnm{Jiahan} \sur{Li}}
\author[5]{\fnm{James H.} \sur{Edgar}}

\author[6]{\fnm{Kenji} \sur{Watanabe}}

\author[7]{\fnm{Takashi} \sur{Taniguchi}}

\author[3]{\fnm{Ni} \sur{Ni}}
\author[8]{\fnm{Su-Yang} \sur{Xu}}

\author[2]{\fnm{Qiong} \sur{Ma}}

\author[1]{\fnm{
Shubhayu} \sur{Chatterjee}} 

\author[1]{\fnm{Jyoti} \sur{Katoch}} 
\author*[1]{\fnm{Simranjeet} \sur{Singh}}\email{simranjs@andrew.cmu.edu }

\affil[1]{\orgdiv{Department of Physics}, \orgname{Carnegie Mellon University}, \orgaddress{ \city{Pittsburgh}, \postcode{15213}, \state{PA}, \country{USA}}}
\affil[2]{\orgdiv{Department of Physics}, \orgname{Boston College}, \orgaddress{ \city{Chestnut Hill}, \postcode{02467}, \state{MA}, \country{USA}}}
%\affil[2]{\orgdiv{Materials Science Division}, \orgname{Argonne National Laboratory}, \orgaddress{ \city{Lemont}, \postcode{60439}, \state{IL}, \country{USA}}}

\affil[3]{\orgdiv{Department of Physics and Astronomy and the California NanoSystems Institute}, \orgname{ University of California}, \orgaddress{ \city{Los Angeles}, \postcode{90095}, \state{CA}, \country{USA}}}

\affil[4]{\orgdiv{Materials and Manufacturing Directorate}, \orgname{ Air Force Research Laboratory}, \orgaddress{ \city{Dayton}, \postcode{45433}, \state{OH}, \country{USA}}}

%\affil[2]{\orgdiv{Department of Physics}, \orgname{Boston College}, \orgaddress{ \city{Chestnut Hill}, \postcode{02467}, \state{MA}, \country{USA}}}

\affil[5]{\orgdiv{Tim Taylor Department of Chemical Engineering}, \orgname{Kansas State University}, \orgaddress{ \city{Manhattan}, \postcode{66506}, \state{KS}, \country{USA}}}

\affil[6]{\orgdiv{Research Center for Electronic and Optical Materials}, \orgname{National Institute for Materials Science}, \orgaddress{ \city{Tsukuba}, \postcode{305-0044}, \country{Japan}}}

\affil[7]{\orgdiv{Research Center for Materials Nanoarchitectonics}, \orgname{National Institute for Materials Science}, \orgaddress{ \city{Tsukuba}, \postcode{305-0044}, \country{Japan}}}

\affil[8]{\orgdiv{Department of Chemistry and Chemical Biology}, \orgname{Harvard University}, \orgaddress{ \city{Cambridge}, \postcode{02138}, \state{MA}, \country{USA}}}

\abstract{Anomalous Hall Effect (AHE) response in magnetic systems is typically proportional to an out-of-plane magnetization component because of the restriction imposed by system symmetries, which demands that the magnetization, applied electric field, and induced Hall current are mutually orthogonal to each other.
Here, we report experimental realization of an unconventional form of AHE in a low-dimensional heterostructure, wherein the Hall response is not only proportional to the out-of-plane magnetization component but also to the in-plane magnetization component. 
By interfacing a low-symmetry topological semimetal (TaIrTe$_4$) with the ferromagnetic insulator (Cr$_2$Ge$_2$Te$_6$), we create a low-dimensional magnetic system, where only one mirror symmetry is preserved.
We show that as long as the magnetization has a finite component in the mirror plane, this last mirror symmetry is broken, allowing the emergence of an AHE signal proportional to in-plane magnetization. 
Our experiments, conducted on multiple devices, reveal a gate-voltage-dependent AHE response, suggesting that the underlying mechanisms responsible for the Hall effect in our system can be tuned via electrostatic gating. 
A minimal microscopic model constrained by the symmetry of the heterostructure shows that both interfacial spin-orbit coupling and time-reversal symmetry breaking via the exchange interaction from magnetization are responsible for the emergence of the in-plane AHE.
Our work highlights the importance of system symmetries and exchange interaction in low-dimensional heterostructures for designing novel and tunable Hall effects in layered quantum systems.
}

%Anomalous Hall Effect (AHE) is conventionally observed in ferromagnets wherein the magnetization, applied electric field, and induced Hall current are mutually orthogonal, a restriction typically imposed by system symmetries. Here, we demonstrate an unconventional AHE in TaIrTe$_4$/CGT heterostructures, where the Hall response is not only proportional to out-of-plane magnetization but also to in-plane magnetization. By interfacing the low-symmetry topological semimetal TaIrTe$_4$ (C$_{2v}$ point group) with the ferromagnetic insulator CGT ($\sim$10 nm), we create a magnetic system in the C$_s$ point group, where only the $M_a$ (\textit{bc}-plane reflection) symmetry is preserved. We show that as long as the magnetization has a finite component in the \textit{bc}-plane of TaIrTe$_4$, this last mirror symmetry is broken, allowing for an in-plane AHE component. Our experiments, conducted on four different devices with bi-layer and tri-layer TaIrTe$_4$, reveal a strong gate-voltage dependence of both in-plane and out-of-plane AHE, suggesting a tunable electronic structure contribution.We propose that the observed AHE arises from exchange interactions in TaIrTe$_4$ induced by CGT magnetization. Our findings highlight how low-symmetry heterostructures enable AHE in unconventional configurations, paving the way for symmetry-engineered Hall effects in layered quantum materials. The experimental results are in good agreement with symmetry analysis.

\maketitle

\section*{Introduction\label{sec:intro}}

The Anomalous Hall effect (AHE) is intrinsically associated with the Berry-phase and has played an important role in probing magnetism and other quantum phenomena in solid-state systems\cite{xiao2010berry,nagaosa2010anomalous}. 
Conventionally, AHE in ferromagnets refers to magnetization-dependent Hall resistance, with the orientation of magnetization, applied electric field, and the Hall current being mutually orthogonal to each other. \cite{karplus1954hall,luttinger1958theory,smit1958spontaneous,berger1970side}. 
Therefore, the AHE and its quantized counterpart (quantum anomalous Hall effect, QAHE), are typically associated with out-of-plane magnetization, i.e., the magnetization is perpendicular to the plane spanned by the applied electric field and Hall current. 
The prerequisite for AHE, i.e., time-reversal ($\mathcal{T}$) symmetry breaking, does not put a constraint on the orientation of the magnetization, instead, the orthogonality restriction in conventional AHE is rooted in material system's rotational and/or mirror symmetries \cite{tan2021unconventional, cao2023plane}.
Unconventional AHE, wherein the Hall resistance is proportional to in-plane magnetization, has been theoretically predicted in magnetic systems with lower crystal symmetry \cite{zhang2011quantized,cao2023plane, wang2024orbital, liu2013plane, tan2021unconventional, liu2018intrinsic, sheoran2025spontaneous, li2024switchable}. 
Experimentally, unconventional Hall effects driven by in-plane magnetization and/or magnetic field have been reported in limited three-dimensional (3D) systems, including the kagome ferromagnet \cite{wang2024orbital,kumar2020anomalous}, magnetic Weyl semimetal \cite{Wang2025In-plane,nakamura2024plane}, topological semimetals where Berry curvature is generated by band topology \cite{liang2018anomalous, galeski2020unconventional, ge2020unconventional}, and magnetic octuple in Fe and Ni films \cite{peng2024observation}. 
In addition, an in-plane AHE has been reported in a heterodimensional superlattice VS$_2$/VS; however, it is primarily attributed to Berry curvature generated by the interplay between spin-orbit coupling (SOC) and an in-plane magnetic field, rather than arising from an in-plane magnetization due to ferromagnetic ordering \cite{zhou2022heterodimensional}.
To date, the experimental realization of an in-plane magnetization-induced AHE in a two-dimensional (2D) system, which can provide a pathway for achieving QAHE in an in-plane magnetic configuration to expand the parameter space for designing dissipationless edge transport in low-dimensional systems \cite{liu2013plane, liu2018intrinsic}, remains notably missing. 
%Unlike widely studied conventional AHE associated with out-of-plane magnetization, experimental studies on in-plane magnetization-induced AHE remain limited.

Here, we report the experimental realization of an unconventional AHE, associated with both out-of-plane and in-plane magnetization, in a 2D topological system. 
To achieve this, we engineer a low-symmetry magnetic heterostructure by combining a low-symmetry topological semimetal, TaIrTe$_4$, with a layered ferromagnetic insulator, Cr$_2$Ge$_2$Te$_6$ (CGT), which is particularly suited for hosting unconventional AHE. 
We demonstrate that the observed Hall response in these heterostructures can be driven solely by the magnetization in CGT, even when it lies entirely in the plane. 
To elucidate the underlying mechanism, we develop a minimal microscopic model constrained by the symmetry of the heterostructure, showing that both interfacial spin-orbit coupling (SOC) and time-reversal symmetry breaking via the exchange interaction from magnetization are essential to the emergence of the in-plane AHE.
Our findings establish the first experimental demonstration of a magnetization-driven in-plane AHE in a 2D heterostructure, and provide theoretical insight into the key ingredients necessary for realizing such unconventional Hall responses, potentially paving the way for their quantum counterparts in future studies.

%AHE in non-collinear antiferromagnets \cite{nakatsuji2015large}. 

%Our results reveal that in-plane magnetization can act as a key tuning parameter for Hall effects in low-symmetry quantum materials. This finding provides new insight into the role of symmetry breaking and Berry curvature engineering in 2D systems. Recently, a fractional quantum AHE was observed in twisted MoTe$_2$ \cite{cai2023signatures}, where strong correlations in moiré bands led to spontaneous ferromagnetism. 

%In contrast, our work presents an alternative approach by introducing exchange interactions into a 2D topological system via interfacing with a ferromagnetic insulator.  

% Our results highlight how interfacial exchange coupling in low-symmetry heterostructures can be leveraged to explore new quantum transport regimes, offering exciting prospects for future investigations into symmetry-protected electronic states and engineered topological phenomena.
%\textcolor{red}{[Write a sentence(s) about how its not been in observed in 2D systems; how in-plane AHE might enable QAHE with in-plane fields]}. \textcolor{blue}{[updated]}

\begin{figure*}[ht]
\centering
\includegraphics[width=1\textwidth]{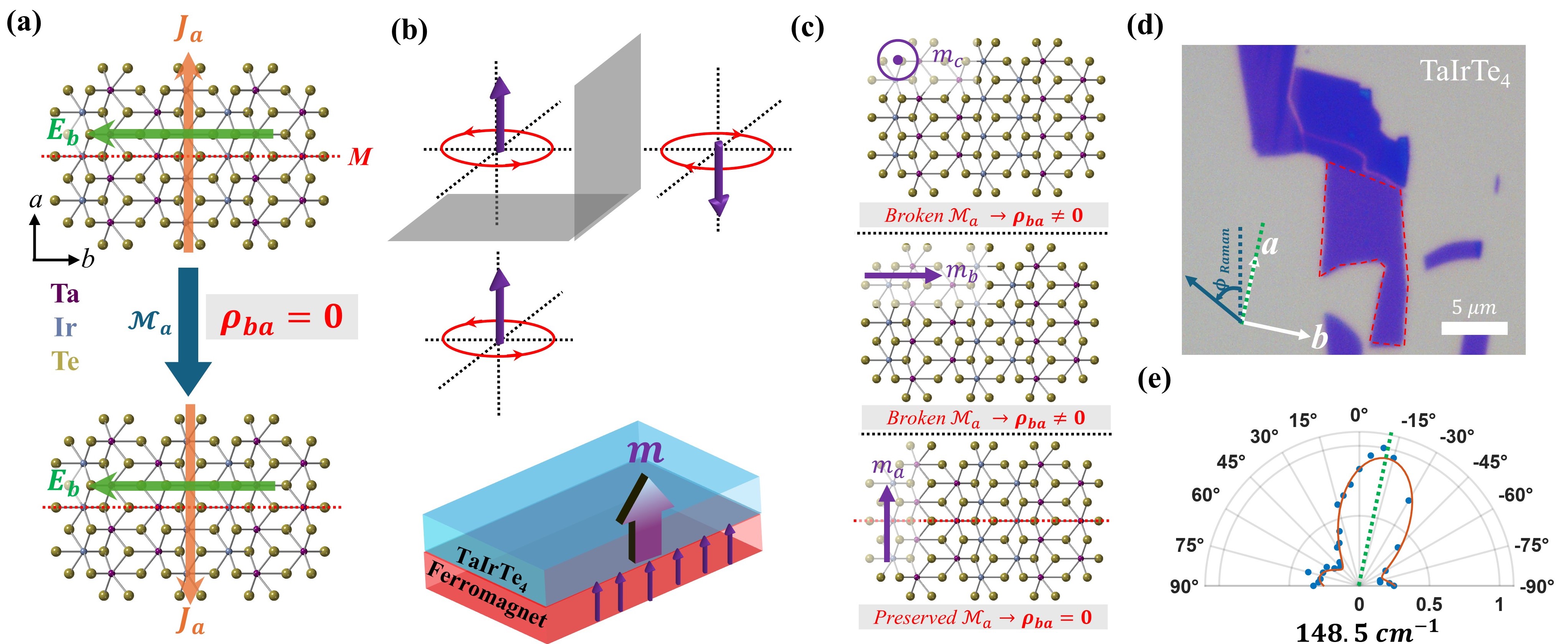}
\caption{\label{Main_Fig1}\textbf{Anomalous Hall effect and crystal symmetry.} 
\textbf{a}, Visualization that AHE is not allowed in TaIrTe$_4$/ CGT heterostructure, in $\mathcal{C}_s$ point group, when magnetic field (magnetization) is absent, as required by the $\mathcal{M}_a$ symmetry. The $\mathcal{M}_a$ symmetry operation reveres $J_a$ but keeps $E_b$ unchanged and, therefore, requires $\rho_{ba}$ to be zero.
\textbf{b}, Upper panel: Illustration showing that magnetization breaks mirror symmetries when their reflection planes are collinear with the magnetization direction, while it preserves those with reflection planes perpendicular to it. 
Lower panel: Demonstration that a low-symmetry magnetic heterostructure can be obtained by coupling a low-symmetry topological semimetal, TaIrTe$_4$, with a ferromagnet.
\textbf{c}, Diagram showing that the AHE is only allowed when there is a finite magnetization component in the \textit{bc}-plane of TaIrTe$_4$, which breaks the $\mathcal{M}_a$ mirror symmetry.
\textbf{d}, Optical image of a bilayer TaIrTe$_4$ flake used in device A, where the chosen region is outlined by the red dashed lines. The polarized Raman measurements were performed with the polarization of the incident laser oriented at an angle $\phi_{\mathrm{Raman}}$ relative to the vertical direction.
\textbf{e}, Angle-dependent polarized Raman spectral intensity at 148.5 cm$^{-1}$ of TaIrTe$_4$ flake in \textbf{d}.}
% Scale bar: 5 $\mu$m
\end{figure*}

\section*{Symmetry Constraints on Anomalous Hall Effect } %Physical picture
% or symmetry controlled anomaloua Hall effect? 

We begin by illustrating the physical origin of unconventional in-plane AHE in 2D materials via symmetry considerations, which motivates our specific heterostructure choice.  
Such symmetry requirements for the emergence of unconventional AHE have been established for general 3D materials \cite{tan2021unconventional} --- all rotational and mirror symmetries must be broken when the magnetization lies in-plane. 
An alternative strategy to engineer AHE in 2D is to build magnetic heterostuctures with sufficiently low symmetry and intrinsic magnetic moments. 
Here, the magnetization is introduced via proximity to a magnetic material, and both the interface and the magnetization serve to reduce the symmetry as desired \cite{liu2020two}, 

Any linear Hall effect, wherein the Hall voltage is proportional to the charge current, can only occur in the absence of a mirror symmetry with the reflection plane parallel to the out-of-plane direction. 
To see this, consider the set-up shown schematically in Fig.~\ref{Main_Fig1}a, where a charge current density $J_a$ is applies along the $a$-axis and a perpendicular electric field $E_b$ is generated due to the Hall effect. 
Under the action of a mirror \( \mathcal{M}_a \), $E_b$ remains unchanged while $J_a$ reverses direction ($J_a \to - J_a$).
However, an electric field induced by any linear Hall effect must reverse sign when the current density is reversed, enforcing the Hall resistivity $\rho_{ba} = E_b/J_a$ to be zero. 
By contrast, if the mirror symmetry \( \mathcal{M}_a \) is broken, either via the interface or the magnetization, then $\rho_{ba}$ is generically non-zero.

Our intuitive picture can be put on a more formal footing by assuming an intrinsic origin of the AHE, wherein the Hall conductivity $\sigma_{ab}$ is set by the integral of the Berry curvature $\Omega(\k)$ over all occupied states in a band with dispersion $\varepsilon(\k)$, i.e., $\sigma_{ab} \propto \int_\k \Omega(\k) \, n_F[\varepsilon(\k)]$ \cite{AHE_RMP2010}.
As long as time-reversal symmetry $\mathcal{T}$ is present, the Berry curvature is odd under $\mathcal{T}$, i.e., $ \Omega(\k) = - \Omega(-\k)$, while the dispersion is even, i.e., $\varepsilon(\k) = \varepsilon(-\k)$, enforcing that $\sigma_{ab}$ vanishes. 
If time-reversal symmetry is broken by magnetization $\mathbf{m}$, the Berry curvature $\Omega(\k)$ can be non-zero.
$\bm{m}$ also plays a crucial role in mirror symmetry: being a pseudovector, it preserves (breaks) mirror symmetry when perpendicular (parallel) to the mirror plane (upper panel, Fig.~\ref{Main_Fig1}b). 
Therefore, if $\mathbf{m}$ is perpendicular to a mirror-plane for the heterostructure, say $\mathbf{m} \parallel \mathbf{a}$, then the mirror symmetry $\mathcal{M}_a$ is preserved.  
Since the Berry curvature is a pseudoscalar in 2D, $\mathcal{M}_a$ requires that the Berry curvature is odd in $k_a$, i.e., $\Omega(k_a,k_b) = - \Omega(-k_a, k_b)$, while the dispersion is even in $k_a$, i.e., $\varepsilon(k_a,k_b)= \varepsilon(-k_a, k_b)$, enforcing that the integral over $k_a$ vanishes and thus $\sigma_{ab} = 0$.
Consequently, one needs to construct a 2D heterostructure with a maximum of a single mirror plane, and arrange for the in-plane magnetization $\mathbf{m}$ to have a component parallel to the said mirror plane, to observe an AHE. 

To substantiate the above argument microscopically, we construct a minimal 2-band model by considering a 2D system with $\mathcal{C}_{2v}$ symmetry, that is reduced to $\mathcal{C}_s$ upon interfacing with a magnetic material 
(details in Methods and Fig. \ref{SI_Fig:Theory}). 
The $\mathcal{C}_{2v}$ point group contains a two-fold rotational symmetry $\mathcal{C}_2$ along the \textit{c}-axis, as well as two mirror symmetries: $\mathcal{M}_a: (a,b,c) \rightarrow (-a,b,c)$, which corresponds to a mirror reflection with respect to the \textit{bc}-plane, and $\mathcal{M}_b: (a,b,c) \rightarrow (a,-b,c)$, corresponding to a mirror reflection with respect to the \textit{ac}-plane.
Therefore, the simplest symmetry-allowed Hamiltonian, to quadratic order in momentum $\bm{k}=(k_a,k_b)$, is given by
\begin{equation}
    H_0 = \frac{k_a^2}{2\tilde{m}_a}+\frac{k_b^2}{2\tilde{m}_b} + g_{ab} k_a s^b + g_{ba} k_b s^a
    \label{Eq. Hamiltonianc2V}
\end{equation}
where $\bm{s}=(s_a,s_b,s_c)$ is the spin, $\tilde{m}_a$ ($\tilde{m}_b$) is the effective mass along $k_a$ ($k_b$), $g_{ab}$ and $g_{ba}$ are SOC constants. 
Without the presence of the magnetic interface, the $\mathcal{C}_2$ and $\mathcal{T}$ symmetries forbid any Berry curvature $\Omega(\k)$, therefore, no intrinsic AHE is allowed.

Next, we consider the presence of a magnetic interface such that the $\mathcal{C}_{2v}$ point group is reduced to the $\mathcal{C}_{s}$ point group, where only one mirror symmetry $\mathcal{M}_a$ is preserved. 
By considering the interfacial SOC and exchange interactions from magnetization, the Hamiltonian from Eq. \ref{Eq. Hamiltonianc2V} is modified to
\begin{align}
    H = H_0 + g_{ac}k_a s^c + \tilde{g}_{ac}k_a^3 s^c - \Delta_{ex}\bm{m}\cdot \bm{s}
    \label{Eq. Hamiltoniancsmagnet}
\end{align}
where $g_{ac}, \tilde{g}_{ac}$ are two interfacial SOC constants that couple $k_a$ and $s_c$, $\bm{m}$ is the magnetization direction in the magnetic material, and $\Delta_{ex}$ is the exchange coupling strength.
Note that $g_{ab}$ and $g_{ba}$ may be renormalized from their bare values in $H_0$ via interfacial SOC. 
When $\bm{m} \parallel \bm{a}$, the mirror-symmetry $\mathcal{M}_a$ is preserved, therefore the intrinsic anomalous Hall conductivity calculated using the Hamiltonian in Eq.~\eqref{Eq. Hamiltoniancsmagnet} vanishes.
However, this constraint is lifted when there is finite magnetization in the \textit{bc}-plane, which breaks the $\mathcal{M}_a$ symmetry.
In this case, we find a nonzero and asymmetric Berry curvature (Fig. \ref{SI_Fig:Theory}e-f), and a resultant non-zero Hall conductance when $\bm{m} \parallel \bm{c}$ and \(\bm{m} \parallel \bm{b}\) (see Fig. \ref{SI_Fig:Theory}g).
Thus, the reduced symmetries of our model allows it to exhibit AHE associated with both out-of-plane and in-plane magnetization.
Notably, the additional SOC terms $g_{ac}, \tilde{g}_{ac}$, arising from broken $\mathcal{C}_2$ symmetry, play a crucial role in enabling an asymmetric Berry curvature distribution $\Omega(\k)$ when \( m_b = \bm{m} \cdot \bm{b} \neq 0 \), highlighting the importance of reduced symmetry for in-plane AHE in 2D heterostructures.

Motivated by our theoretical understanding, we purposefully construct a magnetic low-symmetry heterostructure that enables an AHE with in-plane magnetization.
Specifically, we choose to interface few-layer TaIrTe$_4$ with a ferromagnet (FM), as illustrated in the lower panel of Fig.~\ref{Main_Fig1}b. 
Bulk TaIrTe$_4$ is a type-II Weyl semimetal candidate that belongs to the $\mathcal{C}_{2v}$ point group and the Pmn2$_1$ space group \cite{koepernik2016tairte,haubold2017experimental}, while its monolayer has been identified as a quantum spin Hall insulator \cite{guo2020quantum,tang2024dual}. 
In the TaIrTe$_4$/FM heterostructure, the interface reduces the symmetry to the $\mathcal{C}_s$ point group, preserving only a single mirror symmetry $\mathcal{M}_a$, corresponding to reflection with respect to the \textit{bc}-plane of TaIrTe$_4$. 
We select CGT as the FM layer due to its soft magnetic properties, which allow the magnetization to be aligned in any direction under a moderate magnetic field (\(< 1\) T) \cite{zhang2016magnetic,gong2017discovery}.
Based on our formal theoretical considerations as well as the minimal microscopic model discussed above, the TaIrTe$_4$/CGT heterostructure should exhibit an AHE not only when the magnetization is out-of-plane, but also when it lies entirely in-plane.

The general form of AHE in TaIrTe\(_4\)/CGT heterostructure is summarized in Fig.~\ref{Main_Fig1}c for three orthogonal magnetization configurations.
When \(\bm{m} \parallel \bm{c}\), all mirror symmetries with reflection planes along the \textit{c}-axis, including \(\mathcal{M}_a\), are broken, allowing an AHE response proportional to the magnetization component \( m_c \). 
This also underscores why out-of-plane AHE is commonly observed in high-symmetry magnetic systems: the presence of multiple mirror symmetries with reflection planes along the \textit{z}-axis, which can be simultaneously broken only by an out-of-plane magnetization.
Since we focus on electric measurements in the \textit{ab}-plane of TaIrTe$_4$, the \textit{c}-axis corresponds to the out-of-plane (\textit{z}-axis), making \( m_z = m_c \). 
%, explaining why it is the most observed AHE in ferromagnets.
When \(\bm{m} \parallel \bm{b}\), the mirror symmetry \(\mathcal{M}_a\) is also broken, leading to a unique, unconventional AHE term proportional to the in-plane magnetization \( m_b \). 
% unveiling a unique AHE, which is dependent on the in-plane magnetization. 
Conversely, when \(\bm{m} \parallel \bm{a}\), the mirror symmetry \(\mathcal{M}_a\) is preserved, forbidding any AHE contribution from \( m_a \) (\(\rho_{ba} = 0\)). 
Thus, the total AHE contribution to the Hall resistance $R_{ba}$ in the TaIrTe\(_4\)/CGT heterostructure can be expressed as:
\begin{equation}
   R_{ba} = \Delta R_{AHE}^{z} m_z + \Delta R_{AHE}^{b} m_b
    \label{Eq. 1}
\end{equation}
where \( \Delta R_{AHE}^{z} \) and \( \Delta R_{AHE}^{b} \) are the AHE resistances associated with \( m_z \) and \( m_b \), respectively.

%Note that the Hall resistance of interest here is antisymmetric with respect to both magnetic field and magnetization, satisfying $R_{ba} = -R_{ab}$ due to the Onsager reciprocal relations~\cite{onsager1931reciprocalII}. 
%The transverse resistance in the \textit{ab}-plane of TaIrTe$_4$ measured under an arbitrary current direction can be expressed as $R_{yx} = \frac{V_y}{I_x} = \frac{\sin(2\theta_I)}{2}(R_{b} - R_{a}) + R_{ba}$, where $I_x$ is the applied current, $V_y$ is the transverse voltage, $\theta_I$ is the angle between the current and the \textit{a}-axis of TaIrTe$_4$, and $R_{a}$ ($R_{b}$) is the longitudinal resistance along the \textit{a}-axis (\textit{b}-axis), with $R_{b} > R_{a}$ due to anisotropic resistivity~\cite{liu2018raman, li2024room}. 
%Consequently, the antisymmetric Hall resistance $R_{ba}$ exhibits rotational invariance and remains independent of the measurement orientation relative to the crystal axes.

%\textcolor{red}{
We conclude this section by noting that Onsager reciprocal relations \cite{onsager1931reciprocalII} constrain the Hall resistance to be antisymmetric with respect to both the applied magnetic field $\bm{B}$ and the magnetization $\bm{m}$ in linear response, satisfying $R_{ba}(\bm{B}, \bm{m}) = R_{ab}(-\bm{B},-\bm{m}) = -R_{ab}(\bm{B}, \bm{m})$. 
Therefore, $R_{ba}$ can be extracted from the transverse resistance in the $ab$-plane of of TaIrTe$_4$ measured under an arbitrary current direction via $R_{yx} = V_y/I_x = \frac{\sin(2\theta_I)}{2}(R_{b} - R_{a}) + R_{ba}$, where $I_x$ is the applied current, $V_y$ is the transverse voltage, $\theta_I$ is the angle between the current and the \textit{a}-axis of TaIrTe$_4$, and $R_{a}$ ($R_{b}$) is the longitudinal resistance along the \textit{a}-axis (\textit{b}-axis), with $R_{b} > R_{a}$ due to anisotropic resistivity~\cite{liu2018raman, li2024room}.
Notably, the antisymmetric Hall resistance $R_{ba} = - R_{ab}$ remains independent of the measurement orientation relative to the crystalline axes.

\section*{Device Fabrication and Characterization\label{sec:devfab}}

\begin{figure*}[ht]
\centering
\includegraphics[width=0.9\textwidth]{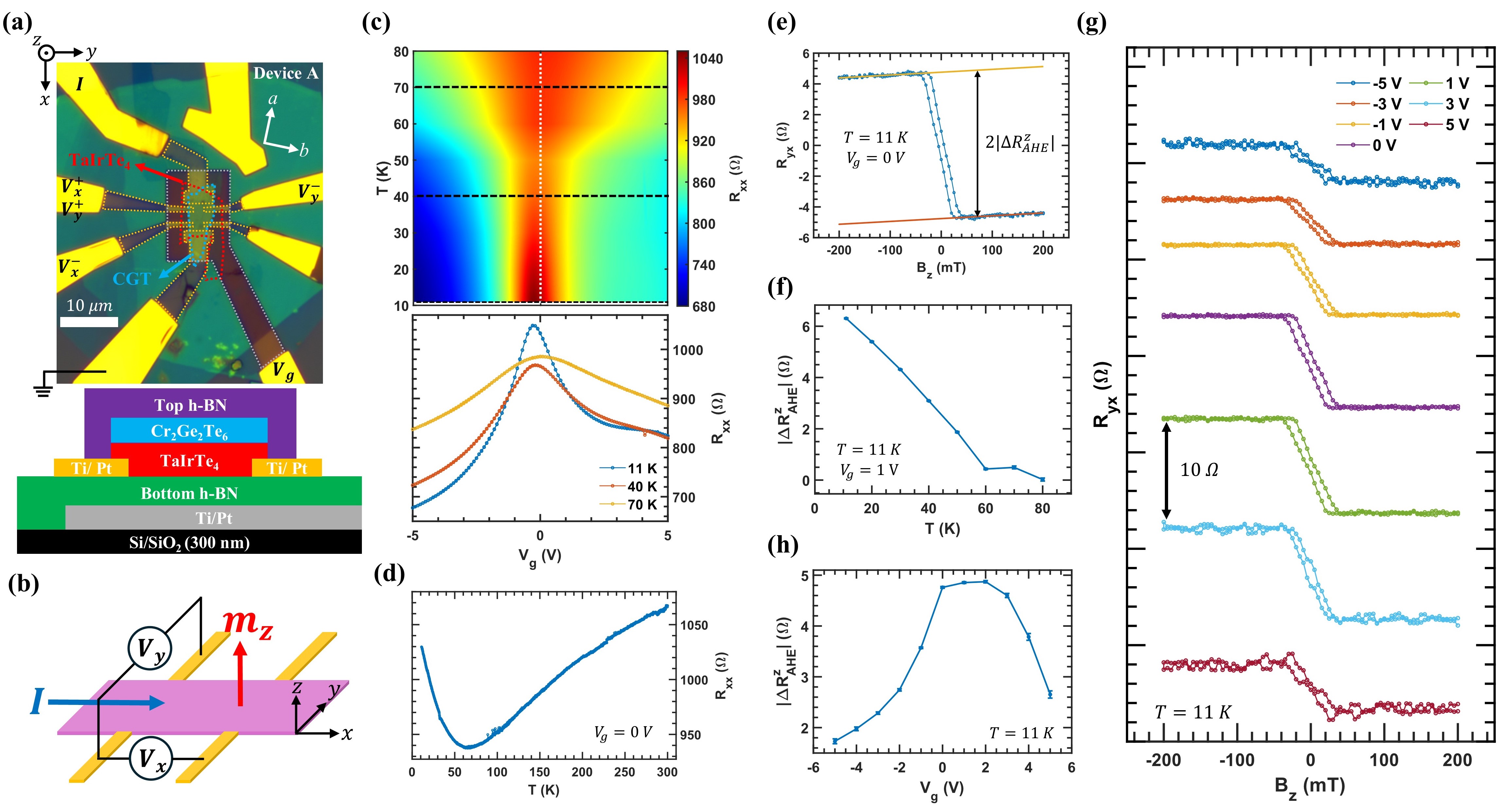}
\caption{\label{Main_Fig2} \textbf{Out-of-plane magnetization induced AHE in TaIrTe$_4$/CGT system.}
\textbf{a}, Upper panel: Optical image of device A with the \textit{a}-axis of TaIrTe$_4$ aligned close to the current electrodes (about $12^\circ$ tilted), with CGT, TaIrTe$_4$, and h-BN flakes outlined and labeled. The Pt electrodes (bottom gate) are outlined with dashed yellow (grey) lines. Lower panel: Schematic of the side view of a typical device.
\textbf{b}, Schematic showing the configuration for the longitudinal and transverse resistance measurements with the magnetization oriented in the out-of-plane direction.
\textbf{c}, Upper panel: Surface map of the measured longitudinal resistance ($R_{xx}$) of Device A as a function of gate voltage ($V_g$), for a range of temperatures (T). Lower panel: The $R_{xx}$ vs $V_g$ of Device A at 11 K, 40 K, and 70 K, as indicated by the black dashed lines in the surface map.
\textbf{d}, The measured $R_{xx}$ of device A as a function of temperature of at $V_g=0~V$.
\textbf{e}, The $R_{yx}$ vs $B_z$ hysteresis loop with the resistance offset removed.
\textbf{f}, The measured $m_z$-dependent AHE resistance, $\Delta R_{AHE}^z$, as a function of temperature at $V_g=1~V$.
\textbf{g}, The $R_{yx}$ vs $B_z$ hysteresis loops at 11 K in a range of $V_g$.
\textbf{h}, $\Delta R_{AHE}^z$ as a function of $V_g$ at 11 K.}
\end{figure*}

For our experiments, we utilize atomically thin flakes of TaIrTe$_4$ (few-layer) and CGT (8-12 nm), and mechanical dry transfer and standard device fabrication techniques were used to assemble van der Waals (vdW) TaIrTe$_4$/CGT heterostructures and their devices \cite{kao2022deterministic,kao2025unconventional} (see Methods). 
We present experimental results from two devices, i.e., namely device A and device B. 
%We will present experimental results from two devices, i.e., namely device A with Hall bar geometry and device B with a 12-point disc geometry. 
To investigate the unconventional AHE in TaIrTe$4$/CGT heterostructures, it is crucial to identify the crystallographic orientation of TaIrTe$_4$ since the in-plane AHE  has a unique in-plane magnetization dependence, which maximizes (vanishes) when $\bm{m}\parallel \bm{b}$  ($\bm{m}\parallel \bm{a}$).
The optical image in Fig.~\ref{Main_Fig1}d shows a bilayer TaIrTe$_4$ flake, where the outlined area is targeted for fabrication of device A. 
We characterized the crystallographic orientation by a combination of parallel cleaving edges and polarized Raman spectroscopy \cite{liu2018raman, tang2024dual,liu2023field}. 
% The Raman spectra are collected by rotating the sample such that the polarization of the incident laser for different angles ($\phi_{Raman}$) relative to the vertical dashed line in the upper panel of Fig.1~\ref{Main_Fig1}d. 
The normalized Raman intensity of TaIrTe$_4$ A$_1$ mode (at $\sim$ 148.5 cm$^{-1}$) at each polarization angle ($\phi_{Raman}$ ) relative to the vertical dashed line in Fig.~\ref{Main_Fig1}d is plotted as a polar plot shown in Fig.~\ref{Main_Fig1}e, depicting a period of 180$^\circ$ and the local intensity maxima occur along \textit{a}- and \textit{b}-axis. 
The A$_1$ mode shows Raman intensity maxima when the polarization is along the \textit{a}-axis of TaIrTe$_4$ \cite{liu2018raman,tang2024dual,liu2023field}, therefore, the maximal intensity indicates the \textit{a}-axis of TaIrTe$_4$ is 12$^\circ$ away from the vertical dashed blue line in Fig.~\ref{Main_Fig1}d.

The optical micrograph of device A, composed of a bilayer TaIrTe$_4$ (Fig. \ref{Main_Fig1}d) and 11.9 nm CGT, is shown in Fig.~\ref{Main_Fig2}a (upper panel) along with a side-view schematic of a typical device (lower panel). 
% The Pt electrodes were patterned into a Hall bar atop an h-BN layer before assembling the final vdW heterostructures. 
% To protect the air-sensitive TaIrTe$_4$ and CGT materials from degradation, the final stack is encapsulated within h-BN flakes.
% Additionally, a Pt gate electrode positioned at the bottom of the device provides electric gating functionality, allowing for tunable control over the electron chemical potential.
The flake thicknesses are confirmed by a combination of optical contrast and atomic force microscopy (AFM) (see Methods and Supplementary information). 
The polarized Raman spectra taken after the device fabrication (Fig. \ref{Main_Fig1}e) confirmed that the \textit{a}-axis of TaIrTe$_4$ in device A is nearly aligned with the current channel of the Hall bar ($\sim$12$^\circ$).
We measure the electronic transport by lock-in amplifier techniques as illustrated in Fig.~\ref{Main_Fig2}b, where a sinusoidal current $I=I_x\sin(\omega t)$ is applied while the in-phase first harmonic longitudinal (transverse) voltage response $V_x$ ($V_y$) is measured.
The current magnitude $I_x=1~\mu A$ is used for all electric transport measurements for both Device A and B. 
In our devices, consisting of an atomically thin TaIrTe$_4$ layer and an insulating magnet, the electric transport in the TaIrTe$_4$ layer is tunable by applying an electrostatic gate voltage ($V_g$), which adjusts the electron chemical potential in the TaIrTe$_4$ layer. 
The measured longitudinal resistance ($R_{xx}=V_{x}/I_x$) for device A as a function of $V_g$ in a range of temperature from 11 K to 80 K is plotted in the upper panel of Fig.~\ref{Main_Fig2}c.
%, wherein a rich unconventional transport with transition near 60 K. 
The line cuts at 11 K, 40 K, and 70 K in the lower panel of Fig.~\ref{Main_Fig2}c show a clear resistance peak at the charge neutrality point (CNP).
In the higher electron-doped (positive $V_g$) and hole-doped (negative $V_g$) regions, device A has metallic behavior, where the resistance monotonically decreases with decreasing temperatures. 
However, near the CNP at $V_g \sim 0~V$, a metal-insulator transition occurs around 60 K, as evident from the temperature dependence of $R_{xx}$ measured at $V_g = 0~V$ shown in Fig.~\ref{Main_Fig2}d.
The measured gate voltage and temperature dependence of $R_{xx}$ are in good agreement with the measured electric transport of bilayer TaIrTe$_4$ \cite{tang2024dual}. 

\section*{Conventional Out-of-plane Anomalous Hall Effect\label{sec:OOPAHE}}

Next, we discuss the experimental observation of the conventional out-of-plane magnetization dependent AHE in device A, and the measurement setup is depicted in Fig.~\ref{Main_Fig2}b. 
The transverse resistances presented in the main text are anti-symmetrized with respect to the magnetic field to isolate the Hall resistance (see Methods).
We measured the Hall resistance as a function of the out-of-plane magnetic field $B_z$ in device A at 11 K, as shown in Fig.~\ref{Main_Fig2}e. 
Evidently, the Hall resistance contains contributions from both the ordinary Hall effect, arising from the applied field $B_z$, and the anomalous Hall effect, associated with the out-of-plane magnetization $m_z$ in the CGT layer.
The saturation observed in $R_{yx}$ at $\pm B_z$ agrees well with the magnetic properties of CGT, i.e., the saturation magnetic field is around $B_z \sim 50$ mT with small coercivity at $\sim10 K$ \cite{zhuo2021manipulating,wang2018electric,gong2017discovery,zhang2016magnetic}. 
The saturated branches at high field ($>100$~mT) were fitted using a linear function and a step function, with the step size defined as $2|\Delta R_{AHE}^z|$ (see Methods).
The resistance component linear in the \textit{B}-field, which is not associated with the magnetization, is removed, allowing us to isolate the anomalous Hall resistance $\Delta R_{AHE}^z$.

This measurement of the AHE hysteresis loop was repeated at various temperatures, and the extracted $\Delta R_{AHE}^z$ is plotted as a function of temperature in Fig. \ref{Main_Fig2}f. 
The AHE signal vanishes near the Curie temperature of CGT ($\sim$ 61 K), indicating that the observed AHE in TaIrTe$_4$ arises from the magnetism in CGT layer \cite{carteaux1995crystallographic,zhang2016magnetic,gong2017discovery}. 
This confirms the prediction of our theoretical model: the combination of interface-induced SOC and time-reversal symmetry breaking via exchange coupling to the magnetization in CGT gives rise to an asymmetric Berry curvature distribution in TaIrTe$_4$, resulting in a nonzero  AHE response.
\cite{kao2025unconventional,lohmann2019probing,lu2013hybrid,mogi2019large,llacsahuanga2022gate,chong2018topological,gupta2022gate,jain2024quantized}.

The observed AHE in device A is gate tunable by changing the electron chemical potential in TaIrTe$_4$ through electrostatic gating. 
% \textcolor{red}{Two comments: (i) It is not quite correct that the Berry curvature is tuned by changing the chemical potential, it is the integral of the Berry curvature that changes because of more filled states. (ii) We can try to see the trend in 2(h) theoretically in our simple model. This may not work because the model is overly simplistic, but worth a try.}
To examine the gate tunability of AHE, we measured the $R_{yx}$ vs $B_z$ hysteresis loops at various $V_g$ and is plotted in Fig.~\ref{Main_Fig2}g.
The extracted $\Delta R_{AHE}^z$ as a function of $V_g$ in Fig.~\ref{Main_Fig2}h shows a peak slightly away from the CNP, while the value is suppressed at higher $\pm V_g$. 
This experimental observation shows that the anomalous Hall resistance is sensitive to variations in the electron chemical potential, consistent with our theoretical expectations (Fig.~\ref{SI_Fig:Theory}g).

\section*{Unconventional In-plane Anomalous Hall Effect\label{sec:IPAHE}}

\begin{figure*}[ht]
\centering
\includegraphics[width=1\textwidth]{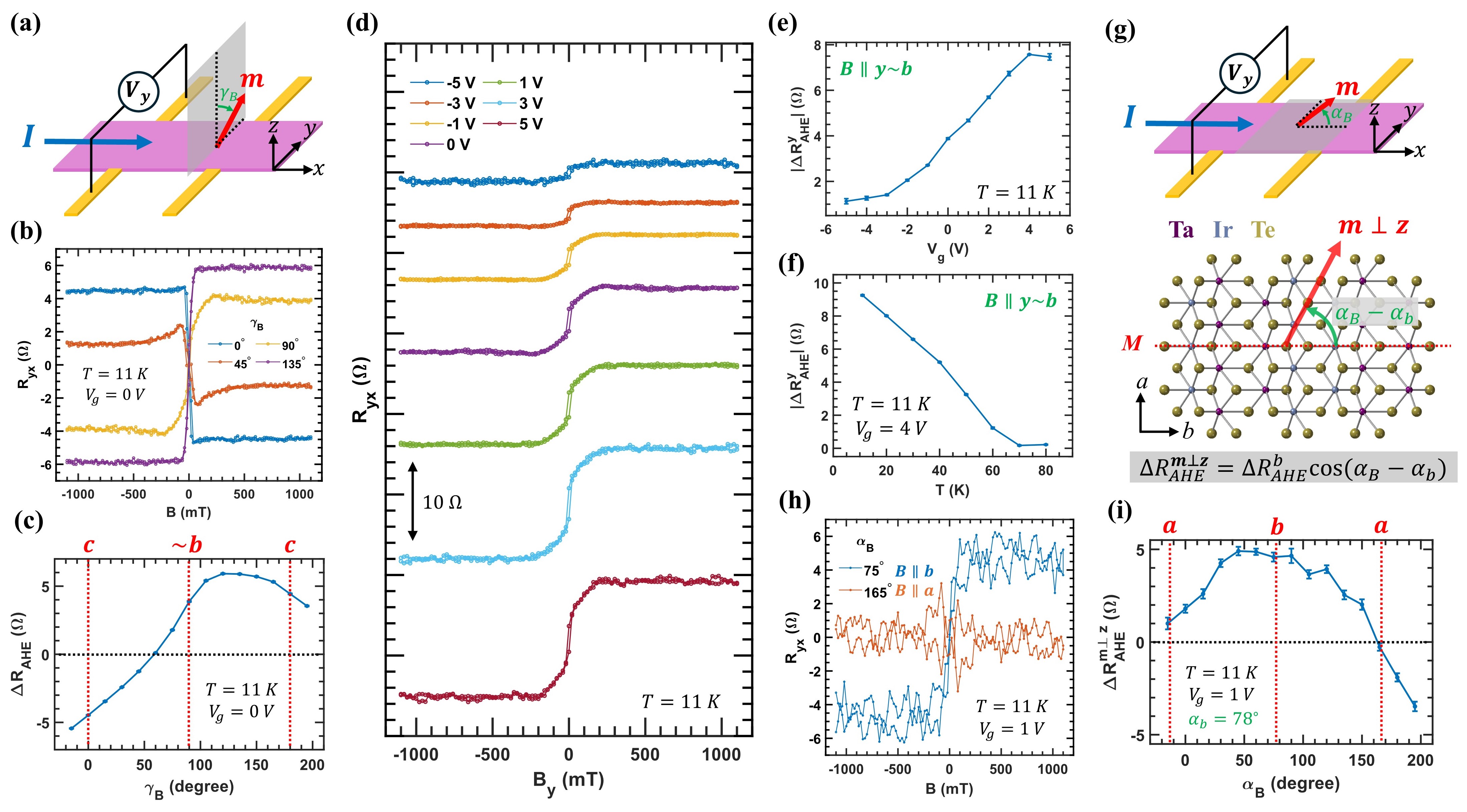}
\caption{\label{Main_Fig3}\textbf{In-plane magnetization induced AHE in TaIrTe$_4$/CGT system.} 
\textbf{a}, Schematic showing the configuration for the Hall resistance measurements with the magnetization oriented at different \textit{zy}-plane rotation angle ($\gamma_B$). 
\textbf{b}, The $R_{yx}$ vs $B$ hysteresis loops with the magnetic field at various \textit{zy}-plane rotation angle $\gamma_B$. Note that the magnetic field is 12$^\circ$ away from \textit{b}-axis of TaIrTe$_4$ when $\bm{B} \parallel \bm{y}$ ($\gamma_B=90^\circ$).
\textbf{c}, The measured AHE resistance, $\Delta R_{AHE}$, as a function of $\gamma_B$. 
\textbf{d}, The $R_{yx}$ vs $B_y$ hysteresis loops at various $V_g$.
\textbf{e}, The measured $m_y$-dependent AHE resistance, $\Delta R_{AHE}^{m_y}$, as a function of $V_g$.
\textbf{f}, $\Delta R_{AHE}^{m_y}$ as a function of temperature.
\textbf{g}, Upper panel: Schematic showing the configuration for the transverse resistance measurements with the magnetization oriented by the magnetic field at different \textit{xy}-plane rotation angle ($\alpha_B$). 
Lower panel: Illustration that the in-plane AHE, $\Delta R_{AHE}^{\bm{m} \perp \bm{z}}$, maximizes (vanishes) when the magnetization is aligned with the \textit{b}-axis (\textit{a}-axis) of TaIrTe$_4$.
\textbf{h}, The $R_{yx}$ vs $B$ hysteresis loops with $\alpha_B$ at $75^\circ$ and $165^\circ$. The magnetic field is aligned with \textit{a}-axis (\textit{b}-axis) of TaIrTe$_4$ at $\alpha_B=75^\circ$ ($\alpha_B=165^\circ$).
\textbf{i}, $\Delta R_{AHE}^{\bm{m} \perp \bm{z}}$ as a function of $\alpha_B$.}
\end{figure*}

\begin{figure*}[ht]
\centering
\includegraphics[width=1\textwidth]{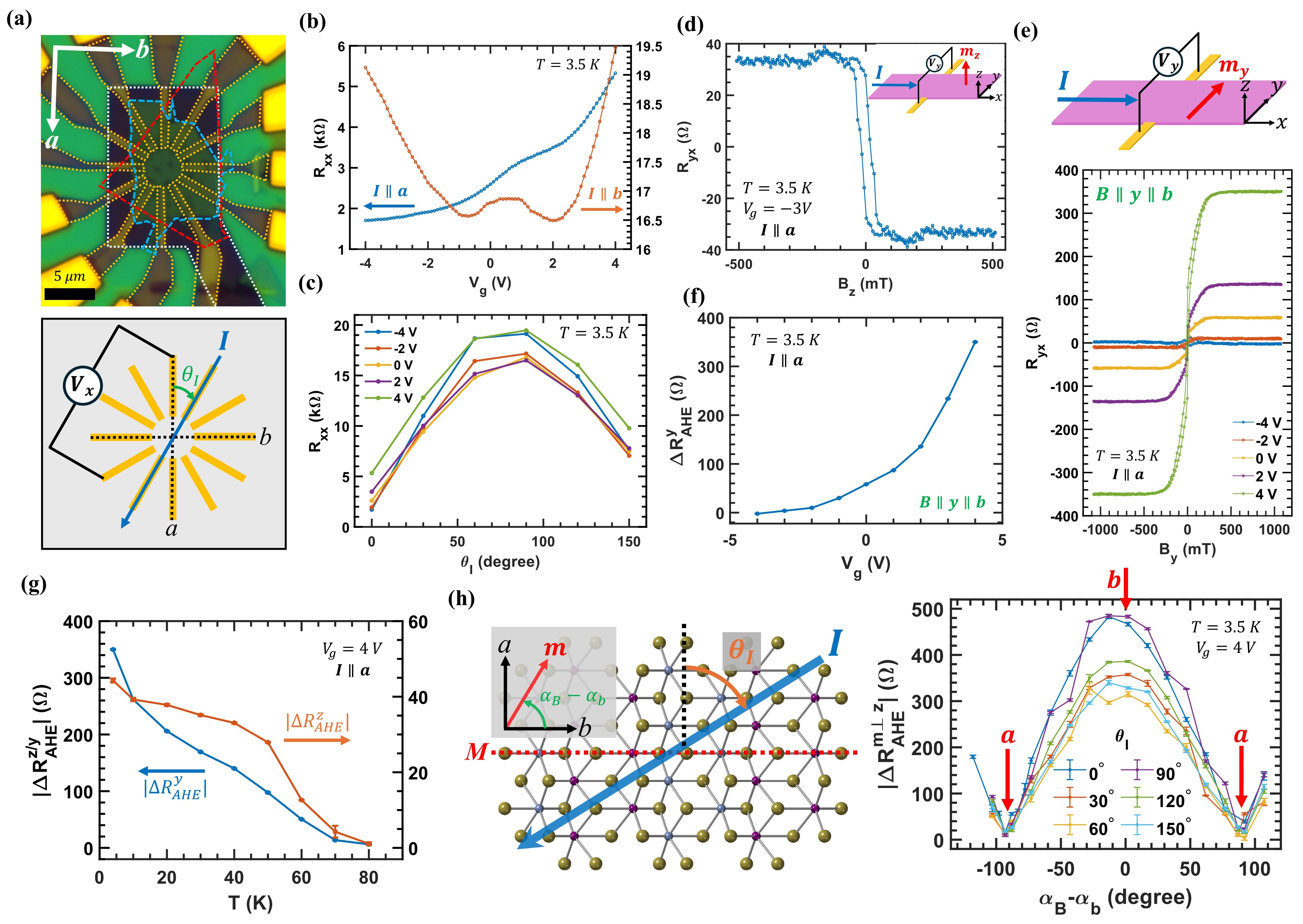}
\caption{\label{Main_Fig4}\textbf{Rotational invariance of in-plane AHE in TaIrTe$_4$/CGT system.}
\textbf{a}, Upper panel: Optical image of device B with the \textit{a}-axis of TaIrTe$_4$ aligned close to one of the current electrodes ($\sim$ $5^\circ$ tilted), with CGT and TaIrTe$_4$ flakes outlined and labeled. The Pt electrodes (bottom gate) are outlined with dashed yellow (grey) lines. 
% Scale bar: 5 $\mu$m. 
Lower panel: Schematic that charge current is applied at different angles $\theta_I$ relative to the \textit{a}-axis of TaIrTe$_4$ while the longitudinal resistance is measured.
\textbf{b}, The $R_{xx}$ as a function of $V_g$ when the charge current is applied along the \textit{a}-axis ($\theta_I=0^\circ$) and \textit{b}-axis ($\theta_I=90^\circ$) of TaIrTe$_4$.
\textbf{c}, The $R_{xx}$ as a function of $\theta_I$ at various $V_g$, where the resistance maximizes (minimizes) when the current is applied along the \textit{b}- (\textit{a}-) axis of TaIrTe$_4$.
\textbf{d}, The $R_{yx}$ vs $B_z$ hysteresis loop with the non-AHE background signals removed. 
Inset: Schematic showing the configuration for the transverse resistance measurements with the magnetization oriented in the out-of-plane direction.
\textbf{e}, Lower panel: The $R_{yx}$ vs $B_y$ hysteresis loops at various $V_g$. With the charge current applied along the \textit{a}-axis of TaIrTe$_4$, the y-direction magnetic field is aligned with the \textit{b}-axis of TaIrTe$_4$. 
Upper panel: Schematic showing the configuration for the transverse resistance measurements with the magnetization oriented in the y-direction.
\textbf{f}, The measured $m_y$-dependent AHE resistance, $\Delta R_{AHE}^{m_y}$, as a function of $V_g$.
\textbf{g}, The $m_y$ ($m_z$) dependent AHE resistance as a function of the temperature at $V_g=4~V$ when $\bm{I}\parallel \bm{a}$.
\textbf{h}, Left panel: An illustration showing that the in-plane AHE magnitude, $|\Delta R_{AHE}^{\bm{m} \perp \bm{z}}|$, is measured when the in-plane magnetization is oriented, by applying an external magnetic field, at an angle of $\alpha_B-\alpha_b$ relative to the \textit{b}-axis of TaIrTe$_4$, while the charge current is applied along various angle $\theta_I$ relative to the \textit{a}-axis of TaIrTe$_4$.
Right panel: The measured $|\Delta R_{AHE}^{\bm{m} \perp \bm{z}}|$ as a function of $\alpha_B-\alpha_b$ when the  the charge current is applied at various $\theta_I$.}
\end{figure*}

To identify the unconventional AHE driven by in-plane magnetization, we leverage the weak magnetic anisotropy of CGT to control the direction of its magnetization $\bm{m}$ using an external magnetic field. 
Since the magnetization of CGT can be easily saturated in any direction with a moderate field ($B > 300~\mathrm{mT}$), this allows us to systematically probe the angular dependence of the AHE.
We performed the $R_{yx}$ vs $B$ hysteresis loop measurements with the \textit{B}-field aligned at different \textit{zy}-plane rotation angles ($\gamma_B$), as illustrated in Fig.~\ref{Main_Fig3}a. 
Based on Eq. \ref{Eq. 1}, $\Delta R_{AHE}(\gamma_B)$ can be written as: 
\begin{equation}
    \Delta R_{AHE}(\gamma_B) = \Delta R_{AHE}^z \cos(\gamma_B)+\Delta R_{AHE}^y \sin(\gamma_B)
    \label{Eq.2}
\end{equation}
where $\Delta R_{AHE}^y= \sin(\alpha_b) \, \Delta R_{AHE}^b$ is the $m_y$-dependent AHE resistance. %\textcolor{red}{$m_y$ or $m_b$?}\textcolor{blue}{IK: this is $m_y$, but in this device, $m_y\sim m_b$ due to oreientation of TaIrTe4 flake and the small misalignment}
The additional factor of $\sin(\alpha_b)$ accounts for the slight misalignment between the TaIrTe$_4$ crystal axis and the electrodes, where $\alpha_b$ is the angle between the \textit{b}-axis and the current channel (\textit{x}-direction).
The polarized Raman results determine $\alpha_b\sim78^\circ$ in device A, however, with such a small misalignment, the measured $\Delta R_{AHE}^y$ is almost equivalent to $\Delta R_{AHE}^b$ with $\sin(\alpha_b)\sim1$.

In Fig.~\ref{Main_Fig3}b, the hysteresis loops measured with different $\gamma_B$ at 0$^\circ$ ($\bm{B}\parallel \bm{z}$), 45$^\circ$, 90$^\circ$ ($\bm{B}\parallel \bm{y}$), and 135$^\circ$ show clear signatures of unconventional in-plane AHE when $\bm{m}\parallel \bm{y}$.
In our measurement setup in device A, we found $\Delta R_{AHE}^z$ ($<0$) and $\Delta R_{AHE}^y$ ($>0$) have opposite signs.
Due to the presence of a large in-plane AHE resistance, the addition (subtraction) of out-of-plane and in-plane AHE results in a larger (smaller) total AHE resistance magnitude at $\gamma_B=135^\circ$ ($\gamma_B=45^\circ$).
Note that the presence of small peaks with $\gamma_B$ at $45^\circ$ and $135^\circ$ is likely due to the PMA in CGT, thus, the magnetization first aligns partially along the out-of-plane direction at smaller \textit{B}-fields when the magnetic field is tilted away from the plane. 
We repeated the hysteresis loop measurements in a range of $\gamma_B$ (from -15$^\circ$ to 195$^\circ$) and obtained the total AHE resistance $\Delta R_{AHE}$ as a function of $\gamma_B$ (Fig.~\ref{Main_Fig3}c).
The measured $\gamma_B$ dependence of $\Delta R_{AHE}$, as described by Eq. \ref{Eq.2}, evidently demonstrates an unconventional in-plane AHE with comparable magnitude to the conventional out-of-plane AHE. 
It is important to emphasize that the observed in-plane AHE cannot arise from a small out-of-plane magnetization component, given the soft magnetic properties of CGT and the fact that the measured total AHE resistance $\Delta R_{AHE}$ would immediately reverse sign across $\gamma_B=90^\circ$ if it were solely associated with $m_z$.

Next, we focused on the in-plane AHE by fixing $\gamma_B = 90^\circ$ such that the magnetic field is aligned along the $\bm{y} \sim \bm{b}$ direction. We measured the temperature dependence of $\Delta R_{AHE}^y$ (Fig.~\ref{Main_Fig3}f), where the in-plane AHE clearly vanishes near the Curie temperature (T$_c$) of CGT, further confirming that the observed unconventional in-plane AHE in TaIrTe$_4$ originates from the magnetism in the adjacent CGT layer.

Third, we investigated the gate voltage tunability of the unconventional in-plane AHE by measuring the $R_{yx}$ versus $B_y$ hysteresis loops at various gate voltages $V_g$ (Fig.~\ref{Main_Fig3}d). Interestingly, the extracted $\Delta R_{AHE}^y$ as a function of $V_g$, shown in Fig.~\ref{Main_Fig3}e, exhibits a strong gate voltage dependence, similar to that observed in $\Delta R_{AHE}^z$. Within the accessible range of $V_g$, we observe a monotonic increase in $\Delta R_{AHE}^y$ with increasing gate voltage.
%allows identification of the suppression of in-plane AHE only in the hole-doped regime (negative $V_g$).

%the low crystal symmetry of TaIrTe$_4$ in combination with the magnetism of CGT.
Finally, we investigated the dependence of the unconventional in-plane AHE on the magnetic orientation within the \textit{ab}-plane of TaIrTe$_4$.
To this end, we perform $R_{yx}$ vs~$B$ hysteresis loop measurements with the in-plane magnetic field aligned at various in-plane rotation angles $\alpha_B$ relative to the current direction (\textit{x}-axis), as illustrated in the upper panel of Fig.~\ref{Main_Fig3}g. As shown in the lower panel of Fig.~\ref{Main_Fig3}g, the symmetry of the TaIrTe$_4$/CGT heterostructure—where only the mirror symmetry with respect to the \textit{bc}-plane is preserved—dictates that the in-plane component of the anomalous Hall effect follows the form:
\begin{equation}
    \Delta R_{AHE}^{\bm{m} \perp \bm{z}}(\alpha_B) = \Delta R_{AHE}^b \cos(\alpha_B - \alpha_{b}), \label{eq.3}
\end{equation}
indicating that the in-plane AHE response reaches its maximum (minimum) when the magnetic field is applied along the $b$-axis ($a$-axis) of TaIrTe$_4$.

%\textcolor{red}{I find this notation a bit confusing, because if $\alpha_B = 0$, the geometric factor is $\cos(\alpha_b)$, but below Eq. (4) the geometric factor was $\sin(\alpha_b)$. Can we use the same notation in both cases? I also find the definition of $\alpha_b$ unclear, what is meant by $xy$-plane rotation angle? It will be much clearer if defined as the angle between two appropriately chosen vector/directions.}
%\textcolor{blue}{revised. I hope this way it is easier to understand. The $\alpha_b$ defined here and in the geometric factor are actually completely identical. The difference between cos and sin here is because the in-plane rotation is defined as an angle relative to the current channel. When we do the out-of-plane to in-plane rotation...in this device A particularly, go through the y-axis (perpendicular to the current channel). }
In Fig.~\ref{Main_Fig3}h, the hysteresis loops measured at different $\alpha_B$, specifically at 75$^\circ$ ($\bm{B} \parallel \bm{a}$) and 165$^\circ$ ($\bm{B} \parallel \bm{b}$), show that the in-plane AHE is finite only when the magnetization has a component aligned with the \textit{b}-axis of TaIrTe$_4$, and is zero when $\bm{m}\parallel \bm{a}$, consistent with the general form of AHE required by the symmetry of TaIrTe$_4$/CGT heterostructures. 
Note that the signal in the \textit{ab}-plane rotation measurement is noisier because the device is affected by cryogenic cooling cycles, necessary for mounting the device in the in-plane rotation position in our magnetotransport setup. 
The in-plane AHE resistance, $\Delta R_{AHE}^{\bm{m}\perp \bm{z}}$, as a function of $\alpha_B$ (Fig.~\ref{Main_Fig3}i) is obtained by measuring the hysteresis loops in a range of $\alpha_B$ from -15$^\circ$ to 195$^\circ$.
In good agreement with our symmetry analysis and the crystal orientation of device A ($\alpha_b \sim 78^\circ$), the unconventional in-plane AHE term $\Delta R_{AHE}^{\bm{m}\perp \bm{z}}$ reaches a maximum when $\bm{m} \parallel \bm{b}$ and diminishes toward zero when $\bm{m} \parallel \bm{a}$.
% [{\color{green}made changes upto here}]

%highlighting the unique characteristics of the unconventional in-plane AHE in the TaIrTe$_4$/CGT heterostructure. 
\section*{Invariance of AHE under Rotations of current orientation}
Next, we present experimental results from device B, shown in Fig.~\ref{Main_Fig4}a (upper panel), which consists of a bilayer TaIrTe$_4$ and CGT. The pre-patterned Pt electrodes in device B are arranged in a 12-point disc geometry, with a 30$^\circ$ interval between adjacent electrodes. 
Device B is designed with two key objectives: first, to confirm that the observed unconventional in-plane AHE is not merely a consequence of the highly anisotropic electronic transport in TaIrTe$_4$; and second, to rigorously test the invariance of the AHE under rotations of the current orientation, thereby providing deeper insight into its underlying symmetry properties.
% The primary goals of device B are to verify the relationship between the observed unconventional in-plane AHE and the crystal symmetry of TaIrTe$_4$, and to provide an alternative method for identifying the crystallographic orientation of TaIrTe$_4$ through anisotropic longitudinal resistance measurements.
The \textit{a}-axis of TaIrTe$_4$ is only $\sim$ 5$^\circ$ tilted from the vertical electrode in device B, verified by the polarized Raman spectroscopy (see Fig. \ref{SI_Fig:DevB_additional}). 
As shown in the lower panel of Fig.~\ref{Main_Fig4}a, the longitudinal resistance is measured when the current is applied at different angles $\theta_I$ relative to the \textit{a}-axis of TaIrTe$_4$.
Due to the anisotropic transport in TaIrTe$_4$, the longitudinal resistance has a form of $R_{xx}=R_a \cos^2\theta_I+R_b \sin^2\theta_I$ \cite{liu2018raman,li2024room}.
The $R_{xx}$ as a function of $V_g$ when $\bm{I}\parallel \bm{a}$ and $\bm{I}\parallel \bm{b}$ in Fig.~\ref{Main_Fig4}b both show a peak near CNP ($V_g\simeq1.5~V$ ), similar to device A. 
The increasing $R_{xx}$ at high electron- and hole-doped regime might be due to the strong electron correlations induced by the formation of van Hove singularities, as previously reported in monolayer TaIrTe$_4$ devices \cite{tang2024dual}.
%Although device B consists of bilayer TaIrTe$_4$, potential degradation of the bottom layer could result in device B exhibiting the electric transport properties of a monolayer TaIrTe$_4$. 
To identify the anisotropic resistance, we plot $R_{xx}$ as a function of $\theta_I$ at various gate voltages in Fig.~\ref{Main_Fig4}c, showing that the longitudinal resistance reaches its maximum (minimum) when the current is applied along the \textit{b}-axis (\textit{a}-axis) of TaIrTe$_4$, consistent with our polarized Raman measurements.

Subsequently, we measured the $R_{yx}$ vs $B_z$ hysteresis loop and observed the conventional out-of-plane AHE, as shown in Fig.\ref{Main_Fig4}d. 
% The peak near zero $B$-field may originate from the topological Hall effect \cite{han2019topological} or the formation of in-plane magnetization due to the gating-induced reduction in perpendicular magnetic anisotropy \cite{verzhbitskiy2020controlling}.
Next, we focus on the in-plane AHE signal in device B.
We first measured the $R_{yx}$ vs $B_y$ hysteresis loops in a range of $V_g$ with $\bm{I} \parallel \bm{a}$ (and consequently $\bm{B} \parallel \bm{b}$), as shown in Fig.\ref{Main_Fig4}e.
The extracted $\Delta R_{AHE}^y \simeq \Delta R_{AHE}^b$ as a function of $V_g$ in device B, plotted in Fig.\ref{Main_Fig4}f, exhibits a $V_g$ dependence similar to the $\Delta R_{AHE}^y$ observed in device A in Fig.\ref{Main_Fig3}e.
%The significantly larger AHE resistance signals in device B are likely due to its transport properties being closer to those of monolayer TaIrTe$_4$. 
%Since the exchange interaction from the adjacent magnetization is primarily an interfacial effect, the observed AHE is expected to diminish in devices composed of thicker TaIrTe$_4$.
The temperature dependence of both out-of-plane (in-plane) AHE resistance magnitude ($\Delta R_{AHE}^{z/y}$) measured at $V_g=4~V$ with $\bm{I} \parallel \bm{a}$ in Fig. \ref{Main_Fig4}g shows that AHE vanishes near the $T_C$ of CGT, consistent with the results from device A. 
Finally, we examined the current direction dependence of the unconventional in-plane AHE by orienting the in-plane magnetization with an in-plane rotation angle of $\alpha_B-\alpha_b$ relative to the \textit{b}-axis of TaIrTe$_4$ when the charge current is applied along different $\theta_I$ relative to the \textit{a}-axis of TaIrTe$_4$, as illustrated by the schematic in the left panel of Fig. \ref{Main_Fig4}h.
As verified by the $|\Delta R_{AHE}^{\bm{m} \perp \bm{z}}|$ as a function of $\alpha_B - \alpha_{b}$ with various $\theta_I$ in the right panel of Fig.~\ref{Main_Fig4}, regardless of the applied current or voltage measurement direction used to measure the transverse resistance, the results consistently show that in-plane AHE magnitude always reaches its maximum at $\alpha_B - \alpha_{b} = 0^\circ$ ($\bm{m} \parallel \bm{b}$) and vanishes at $\alpha_B - \alpha_{b} = \pm90^\circ$ ($\bm{m} \parallel \bm{a}$).
%\textcolor{red}{Do we need an arbitrary $n$? The total angle can only have a range of 360 degrees, which should limit what values n can take(?)}
%\textcolor{blue}{revised}
This observation is again in excellent agreement with the unique form of in-plane AHE in the TaIrTe$_4$/CGT system as required by the symmetry of $\mathcal{C}_s$ point group.
In addition, our experimental findings further corroborate that the observed unconventional in-plane AHE is independent of the current orientation relative to the crystalline axes, as expected for an antisymmetric Hall effect. 
%\textcolor{red}{Once again, I would replace 'shows rotational invariance' by 'is independent of the current orientation relative to the crystalline axes', but if this is standard terminology feel free to keep it.}
%\textcolor{blue}{revised, I think this is also fine.}

\section*{Conclusion} \label{sec:conclusion}
Our work establishes a new paradigm for realizing unconventional forms of AHE in low-dimensional magnetic heterostructures. 
By interfacing a low-symmetry topological semimetal (TaIrTe$_4$) with a ferromagnetic insulator (CGT), we realized a low-dimensional material platform in which the Berry curvature distribution can be tuned by the reduced system symmetry, interface-induced SOC, and magnetic exchange couplings, allowing for the emergence of an unconventional AHE, i.e., an AHE signal proportional to both in-plane and out-of-plane magnetization.
Through systematic experiments on multiple devices, we demonstrated that the AHE in TaIrTe$_4$/CGT presents when the CGT magnetization has a finite component along the remaining mirror ($bc$) plane of TaIrTe$_4$.
Furthermore, we highlighted that the symmetry-controlled in-plane AHE reaches a maximum when the magnetization is aligned along the high-symmetry $b$-axis of TaIrTe$_4$, and vanishes when the magnetization is strictly along the $a$-axis, where the preserved mirror symmetry prohibits a Hall response, consistent with our theoretical symmetry analysis.
Moreover, we observed a clear gate voltage dependence in both AHE components, indicating that the anomalous Hall effect in this 2D magnetic system is highly tunable via electrostatic control of the electron chemical potential.
%\textcolor{red}{We did not find evidence for tuning of Berry curvature}
%\textcolor{blue}{revised}
These findings provide a direct experimental manifestation of symmetry-engineered transport properties in low-symmetry quantum materials.
Beyond confirming the role of symmetry in governing AHE, our results reveal the potential for designing tunable Hall effects by controlling crystal symmetry and exchange interactions at interfaces. 
Our work opens the possibilities of future investigations into other low-symmetry topological materials to engineer novel quantum transport for next-generation spintronic and electronic devices.
%In contrast to conventional AHE (proportional to out-of-plane magnetization), orientation of magnetization, applied electric field, and the Hall current being mutually orthogonal to each other

%which is typically constrained by orthogonal magnetization configurations
%Our study establishes a new paradigm to realize new form of AHE in low-symmetry magnetic heterostructures. By interfacing a topological semimetal (TaIrTe$_4$) with a ferromagnetic insulator CGT, we realized a system in which mirror symmetry breaking allows for an unconventional in-plane AHE component. 
%This stands in contrast to conventional AHE, which is typically constrained by orthogonal magnetization configurations. 

\bibliography{sn-bibliography}   % Main text references

\section*{Methods}\label{Sec:Methods}

\paragraph{Device Fabrication}
TaIrTe$_4$ and hexagonal boron nitride (h-BN) crystals were prepared using previously published procedures \cite{cai2019observation,liu2018single}.
Cr$_2$Ge$_2$Te$_6$ (CGT) single crystals were purchased from HQ Graphene. 
Mechanical exfoliation of TaIrTe4$_{4}$, Cr$_2$Ge$_2$Te$_6$ (CGT), h-BN, and graphite was performed on separate silicon wafers with 300 nm of SiO$_{2}$ inside a glovebox filled with Ar gas. 
Flakes were selected through optical investigation through a microscope. 
TaIrTe4$_{4}$ flakes that have well-defined and straight edges were used because the \textit{a}-axis tends to be along them. 

The heterostructure was fabricated using a custom transfer tool inside a glovebox filled with Ar gas. 
A transfer slide consisting of a polydimethylsiloxane (PDMS) slab and a thin film of polycarbonate (PC) was used for transfer 2D flakes. 
The electrodes for electric connection are patterned using electron beam lithography (EBL) and electron beam deposition with a PMMA/MMA bilayer resist.
The bottom h-BN is transferred on top of a bottom Pt or graphite gate prepared on the Si/SiO$_2$ substrate, and Pt electrodes composed of Ti (2 nm)/Pt (6 nm) were patterned on top of the h-BN.
The Pt electrodes were then connected by Cr (5 nm)/Au (55 nm) electrodes for wire bonding pads.
The Pt electrodes and bottom h-BN were cleaned by atomic force microscopy in contact mode using $\mu$masch HQ:NSC15/Al BS tips and ultra-high vacuum annealing at 200$^\circ$C for 4 hours before the heterostructure of TaIrTe$_4$/CGT/h-BN was transferred to the electrodes to ensure the interface quality. 
For devices with a top gate (Device D), an additional graphite flake is picked up in the beginning to contact the pre-patterned Pt electrodes.

\paragraph{Device Characterization}
To determine the thickness and the crystallographic orientation of the TaIrTe$_4$/CGT devices, we utilize AFM and polarized Raman spectroscopy. 
All the devices were examined by AFM in the tapping mode using $\mu$masch HQ: NSC15/ Al BS tip to investigate the thickness of flakes used in each device. 
The TaIrTe$_4$ thickness is characterized by a combination of optical contrast and AFM. 
The crystallographic orientation of TaIrTe$_4$ in device A, B, and D were confirmed by angle-dependent Polarized Raman measurements. 
We focused on the Raman shift peak of TaIrTe4$_4$ $A_1$ mode at around 148.5 cm$^{-1}$ to determine the crystallographic orientations of the TaIrTe$_4$ flake \cite{liu2018raman, tang2024dual,liu2023field}. 
The Raman intensity of $A_1$ mode is maximized when laser polarization is parallel to the \textit{a}-axis of TaIrTe$_4$. 

\paragraph{Electrical measurements}

Electrical measurements were performed at variable temperatures under high vacuum ($<10^{-5}$ mtorr) conditions.
An electromagnet was rotated such that the magnetic field could be rotated in \textit{xy} or \textit{zy} plane depending on the mounting orientation of the device. 
For device A and B, a Keithley 6221 current source is used to apply a sinusoidal AC current of 1 $\mu A$ at 111.37 Hz, while an SR830 Lock-In Amplifier (Signal Recovery 7265 DSP Lock-in Amplifier) is used to measure the first harmonic transverse (longitudinal) voltage.
For device C and D, a Keithley 6221 current source is used to apply a positive and negative DC current in the range of $1-10~\mu A$, while two Keithley 2182A nanovoltmeters are used for DC measurements for transverse and longitudinal resistances. 
The results obtained by DC methods are symmetrized with respect to the current polarity.
Keithley 2400 and 2450 source meters are used for applying gate voltage to the device. 

The Hall effects are antisymmetric to the B-field, therefore the measured transverse resistance is first anti-symmetrized with respect to B-field to better quantify the observed unconventional AHE. 
This is achieved by measuring the full magnetic hysteresis loop ($-B \rightarrow +B$ and $+B \rightarrow -B$), calculating the transverse Hall resistance by $R_{yx}(B)=\frac{1}{2}[R_{yx,raw}(B)-R_{yx,raw}(-B)]$ between the forward and backward loops. 
Some raw transverse resistance $R_{yx}$ vs B-field results used in the main text are shown in Fig.\ref{SI_Fig:symmetrization}. 

\paragraph{Minimal microscopic two-band model}

We propose an intrinsic origin of the observed AHE, namely, a Hall conductivity induced by Berry curvature of the system.
To this end, we write down a minimal effective Hamiltonian that is constrained by the relevant symmetries, and analyze the resultant Hall conductivity.
Since the symmetries of few-layered TaIrTe$_4$ are described by the $C_{2v}$ point group, which is composed of $\{e, C_2, M_a, M_b \}$, plus the time-reversal operator $\mathcal{T}$, we list the actions of the group elements on momentum and spin below:
\begin{align*} 
\mathcal{C}_{2}:& (k_a, k_b) \to (-k_a, -k_b), \, (s^a, s^b, s^c) \to (-s^a, - s^b, s^c) \nonumber \\
\mathcal{M}_a :& (k_a, k_b) \to (-k_a, k_b), \,
(s^a, s^b, s^c) \to (s^a, - s^b, -s^c) \nonumber \\
\mathcal{M}_b :& (k_a, k_b) \to (k_a, -k_b), \,
(s^a, s^b, s^c) \to (-s^a, s^b, -s^c) \nonumber \\
\mathcal{T}~:& (k_a, k_b) \to (-k_a, -k_b), \, (s^a, s^b, s^c) \to (-s^a, - s^b, -s^c)
\end{align*}
Note that $\mathcal{M}_a \mathcal{M}_b = \mathcal{C}_2$, so we only need to consider one of the mirrors, say $\mathcal{M}_a$. 

Given these symmetries, the most general two-band Hamiltonian that can be written in an expansion in momentum (upto second order) is 
\begin{equation*}
H_{0} = \frac{k_a^2}{2\tilde{m}_a} + \frac{k_b^2}{2\tilde{m}_b} + g_{ab} k_a s^b + g_{ba} k_b s^a
\end{equation*} 
If $g_{ab}= - g_{ba}$, then the spin-orbit coupling (SOC) term takes the form $(\bm{k} \times \bm{s}) \cdot \hat{\bm{z}}$, which is the conventional Rashba SOC. 
This constraint is typically enforced by a larger rotational symmetry about the \textit{c}-axis, but $g_{ab}$ and $g_{ba}$ can be independent if only $\mathcal{C}_2$ is present.
The Berry curvature changes sign under the action of $\mathcal{C}_2 \mathcal{T}$, which is a preserved symmetry of TaIrTe$_4$, so the Berry curvature must be zero and therefore any intrinsic contribution to AHE is forbidden. 
One can simply add magnetization in the \textit{c}-direction to break $\mathcal{C}_{2} \mathcal{T}$ and the mirror symmetry and get a non-zero anomalous Hall conductivity. 
However, an in-plane magnetization by itself is odd under both $C_2$ and $\mathcal{T}$, and consequently does not break $C_{2} \mathcal{T}$.
Therefore no in-plane AHE is allowed when we have $C_{2}\mathcal{T}$, and the symmetry of the system needs to be lowered. 

To achieve the desired lower symmetry, a material with point group $\mathcal{C}_{2v}$, such as TaIrTe$_4$, is interfaced with a magnetic material like CGT to break $\mathcal{C}_2$. 
In this case, the point group is reduced to $\mathcal{C}_s = \{e, \mathcal{M}_a\}$ even in the absence of magnetization $\bm{m}$ in the magnetic material. 
Now, additional combinations of momentum and spin, namely, $k_a s^c$ and $k_a^3s^c$ are symmetry-allowed in the Hamiltonian (the inclusion of the cubic term is required to avoid a fine-tuned limit, as we will explain later).
Furthermore, we include the exchange interactions from the magnetization, so the Hamiltonian becomes
\begin{equation*}
\begin{split}
H = H_0 + g_{ac} k_a s^c + \tilde{g}_{ac} k_a^3 s^c - \Delta_{ex}\bm{m}\cdot\bm{s}
\end{split}
\end{equation*}
where the time-reversal symmetry can be broken by the magnetization to allow for AHE. 
Next, we compute the anomalous Hall conductance when $\bm{m}$ points along each of the crystalline axes.  
For simplicity, we will set the effective masses $\tilde{m}_a = \tilde{m}_b = \tilde{m}$ for the rest of the discussion.

When $\bm{m}$ is in the \textit{b}-direction, the mirror symmetry $\mathcal{M}_a$ is broken.
In this case, the dispersions of the two bands are given by 
% \begin{align}
%     \varepsilon_{1,2}(\mathbf{k}) &=  \frac{k^2}{2\tilde{m}}\nonumber \\ 
%     &\pm \sqrt{(g_{ab}k_a - m)^2 + (g_{ba} k_b)^2 + (g_{ac}k_a + \tilde{g}_{ac} k_a^3)^2},
% \end{align}
% and the Berry curvature of the two bands are \cite{BerryPhase_RMP2010}
% \begin{equation}
%     \Omega_{1,2}(\k) = \pm \frac{m(g_{ba}g_{ac} + 3g_{ba}\tilde{g}_{ac}k_a^2) - 2g_{ab}g_{ba}\tilde{g}_{ac} k_a^3}{2[( g_{ab} k_a - m)^2 + (g_{ba}k_b)^2 + (g_{ac}k_a + \tilde{g}_{ac}k_a^3)^2]^\frac{3}{2}}.
% \end{equation}

\begin{align*}
    \varepsilon_{1,2}(\mathbf{k}) = \frac{k^2}{2\tilde{m}}
    \pm \sqrt{(E_{ab} - m)^2 + E_{ba}^2 + E_{ac}^2}
\end{align*}
and the Berry curvatures of the two bands \cite{xiao2010berry} are 
\begin{equation*}
    \Omega_{1,2}(\k) = \pm \frac{m(g_{ba}g_{ac} + 3g_{ba}\tilde{g}_{ac}k_a^2) - 2g_{ab}g_{ba}\tilde{g}_{ac} k_a^3}{2[( E_{ab} - m)^2 + E_{ba}^2 + E_{ac}^2]^\frac{3}{2}}
\end{equation*}
where $E_{ab}=g_{ab}k_a$, $E_{ba}=g_{ba}k_b$, and $E_{ac}=g_{ac}k_a+\tilde{g}_{ac}k_a^3$.
Note that the Berry curvature is neither even nor odd under $k_a \to -k_a$, and this asymmetry is responsible for a non-zero anomalous Hall conductance.
Specifically, the anomalous Hall conductivity \cite{AHE_RMP2010} is then given by 
\begin{equation*}
    \sigma_{ab} = \sum \limits_{n = 1,2} -\frac{e^2}{\hbar} \iint n_F(\varepsilon_n(\mathbf{k})) \Omega_{n}(\mathbf{k}) \frac{dk_a dk_b}{(2\pi)^2}
    \label{AHC}
\end{equation*} where $n_F$ is the Fermi-Dirac distribution. 
In the $T \to 0$ limit, the integral runs over all the occupied states, chosen to be $[-0.01, 0.01]nm^{-1}$ in both directions to cover all the space where the Berry curvature is significant. 
The resultant Hall conductance for $\bm{m} \parallel \bm{b}$ is shown in Fig.~\ref{SI_Fig:Theory}(g).

We can apply this approach to calculate the anomalous Hall response for the magnetization vector pointing in other directions too. 
In particular, let us explicitly see why the anomalous Hall conductance must be zero when the magnetization $\bm{m}$ is along the \textit{a}-axis and the system preserves $\mathcal{M}_a$ symmetry. 
In this limit, the band dispersions are
\begin{align*}
    \varepsilon_{1,2}(\mathbf{k}) = \frac{k^2}{2\tilde{m}} \nonumber  
    \pm \sqrt{E_{ab}^2 + (E_{ba}-m)^2 + E_{ac}^2}
\end{align*}
while the Berry curvatures take the form
\begin{equation*}
    \Omega_{1,2}(\k) = \pm \frac{g_{ab}g_{ba}\tilde{g}_{ac}k_a^3}{[E_{ab}^2 + (E_{ba}-m)^2 + E_{ac}^2]^\frac{3}{2}}
\end{equation*}
We note that the Berry curvatures $\Omega_{1,2}(\k)$ would vanish if we did not include the cubic term $\tilde{g}_{ac} k_a^3 s^c$, making this example too fine-tuned. 
% In addition, $\Gamma$ would be a single Dirac point, which violates the Nielsen-Ninomiya theorem\cite{NIELSEN198120, NIELSEN1981219, NIELSEN1981173}. But now $\tilde{g}_{ac} k_a^3 s^c$ breaks the chiral symmetry $\frac{1}{\sqrt{2}} (s^b - s^c)$, so we know that $\Gamma$ is not a Dirac point with the $\mathcal{C}_{2} \mathcal{T}$ symmetry broken already\cite{PhysRevB.106.045126}.
% Now we apply $\mathcal{M}_a$ to Eq.\ref{AHC}. Note that Berry curvature is a pseudo-scalar in 2D, so $\Omega_\pm(k_a, k_b) = - \Omega_{\pm}(-k_a, k_b)$. Meanwhile, the preserved $\mathcal{M}_a$ symmetry also enforces $E(k_a, k_b) = E(-k_a, k_b)$, so after $\mathcal{M}_a$

From our computations, we can explicitly see that the dispersions are even under $k_a \to -k_a$, i.e., $\varepsilon_{1,2}(k_a, k_b) = \varepsilon_{1,2}(-k_a, k_b)$. while the Berry curvature is odd under $k_a \to - k_a$, i.e., $\Omega_{1,2}(k_a, k_b) = - \Omega_{1,2}(-k_a, k_b)$, as expected for a pseudoscalar in 2D.
This is a consequence of preserving the mirror symmetry $\mathcal{M}_a$, and the resultant Hall conductance can be shown to vanish by acting $\mathcal{M}_a$ on the integrand, as follows:
\begin{align*}
\sigma_{ab} & = - \frac{e^2}{\hbar} \sum_{n = 1,2} \iint  n_F[\varepsilon_n(-k_a, k_b)] \Omega_n(-k_a, k_b) \frac{dk_a dk_b}{(2\pi)^2}\nonumber \\
& = -\frac{e^2}{\hbar} \sum_{n = 1,2} \iint  n_F[\varepsilon_n(k_a, k_b)]\left(- \Omega_n(k_a, k_b) \right) \frac{dk_a dk_b}{(2\pi)^2}\nonumber \\
& = -\sigma_{ab} \implies \sigma_{ab} = 0
\end{align*}

For completeness, we also write down the expressions of band dispersions and Berry curvatures for magnetization $\bm{m}$ along the $c$-axis:
\begin{align*}
    \varepsilon_{1,2}(\mathbf{k}) = \frac{k^2}{2\tilde{m}} \nonumber 
    \pm \sqrt{E_{ab}^2 + E_{ba}^2 + (E_{ac} - m)^2},
\end{align*}
and
\begin{equation*}
    \Omega_{1,2}(\k) = \mp \frac{g_{ab}g_{ba}(m+2\tilde{g}_{ac}k_a^3)}{2[E_{ab}^2 + E_{ba}^2 + (E_{ac}- m)^2]^\frac{3}{2}}.
\end{equation*}
% Note that in this case there is also the usual Hall effect due to Lorentz force in presence of a non-zero external magnetic field $\bm{B} \parallel \bm{c}$, but it is not of interest to us.

We summarize our results in Fig.~\ref{SI_Fig:Theory}. With reasonable parameters, the model yields a finite and physically consistent anomalous Hall conductivity. 
Theoretical results are shown for both in-plane and out-of-plane magnetization configurations, offering a qualitative explanation for the emergence of unconventional AHE in magnetic systems with $\mathcal{C}_s$ symmetry, such as TaIrTe$_4$/CGT heterostructures. 
The key ingredients in our model are the spin-orbit coupling (SOC) terms $k_a s^c$ and $k_a^3 s^c$ that are symmetry-allowed (even though we do not comment on their microscopic origin), along with the exchange coupling $\Delta_{ex}\bm{m} \cdot \bm{s}$. 
If the observed AHE is primarily driven by Berry curvature, these terms offer a natural and minimal explanation. We emphasize that this model is not intended to quantitatively match experimental data, but rather to propose a simple, symmetry-based, microscopic picture that captures the essential physics of our experimental observations.

\section*{Acknowledgments}

S.S. acknowledges the financial support from U.S. Office of Naval Research under Award No. N00014-23-1-2751, National Science Foundation (NSF) through Grants No. ECCS-2208057 and DMR-2210510, and from the Center for Emergent Materials at The Ohio State University, a National Science Foundation (NSF) MRSEC, through Award No. DMR-2011876. 
S.S. also acknowledges financial support from NSF-CAREER Award through Grant No. ECCS-2339723. 
J.K. acknowledges the financial support from U.S. Office of Naval Research under Award No. N00014-23-1-2751, the Center for Emergent Materials at The Ohio State University, an NSF MRSEC, through Award No. DMR-2011876, and the U.S. Department Office of Science, Office of Basic Sciences, of the U.S. Department of Energy through award No. DE-SC0020323 (for device fabrication). 
J.K. also acknowledges financial support from NSF-CAREER Award under Grant No. DMR-2339309. 
The single crystal growth and characterization of TaIrTe$_4$ at UCLA were supported by the U.S. Department of Energy (DOE), Office of Science, Office of Basic Energy Sciences under Award Number DE-SC0021117.
J.H.E. acknowledges the support for hBN crystal growth from the U. S. Office of Naval Research under award number N00014-22-1-2582. 
K.W. and T.T. acknowledge support from 
the JSPS KAKENHI (Grant Numbers 21H05233 and 23H02052), the CREST (JPMJCR24A5), JST and World Premier International Research Center Initiative (WPI), MEXT, Japan.
We acknowledge Archibald J. Williams for providing the schematic of TaIrTe$_4$ crystal structure used in the manuscript Figures. The authors also thank Ran Cheng and Junyu Tang for insightful discussions.

\section*{Author Contributions}
S.S. and J.K. supervised the research. 
I.K. and R.K.B. prepared the devices, performed measurements, and analyzed the data with assistance of Z.C., S.S., A.T., and M.C.   
J.T., Q.M., and S.Y.X provided the support for sample and device preparation.
S.Z. and S.C. provided the theoretical support.
R.R. carried out polarized Raman measurements. 
T.Q. and N.N. grew the bulk crystals of TaIrTe$_4$. 
J.L., J. H. E., K.W., and T.T. provided the bulk h-BN crystals. All authors contributed to write the manuscript.

\section*{Competing interests}
The authors declare no competing interests. 
% \clearpage
% \bibliography{method-bibliography}   % Main text references

%%%%%%%%%%%%%%%% END OF MAIN TEXT %%%%%%%%%%%%%%%

\clearpage
\newpage

%%%%%%%%%%%%%%%% START OF SUPPLEMENT %%%%%%%%%%%%%%%
% \begin{appendices}

% Figures, tables, equations and pages in the supplement are numbered S1, S2 etc.
% \renewcommand{\thefigure}{\arabic{figure}} % Reset figure numbering
% \renewcommand{\figurename}{Extended Data Figure} % Change "Figure" to "Extended Data Figure"
\setcounter{figure}{0}  % Reset figure counter

\setcounter{table}{0}  % Reset figure counter

\setcounter{equation}{0}  % Reset figure counter

\renewcommand{\thefigure}{S\arabic{figure}}
\renewcommand{\thetable}{S\arabic{table}}
\renewcommand{\theequation}{S\arabic{equation}}
\renewcommand{\thepage}{S\arabic{page}}

\setcounter{page}{1} % not 0 as \newpage already started a supplementary page
% % References continue the numbering from the main text.

\onecolumn
% \section*{Extended Data}\label{Appe:Extended Data}
\begin{center}
\section*{Supplementary Information}\label{Appe:supplementary inforamtion}
\end{center}

\begin{table}[hbt!]
\centering
\caption{The TaIrTe$_4$ alignment, electrode geometry, and thickness parameters of TaIrTe4$_4$, CGT, thBN, and bhBN flakes in TaIrTe$_4$/CGT devices.}
\begin{tabularx}{\textwidth}{ 
   >{\centering\arraybackslash}X
   >{\centering\arraybackslash}X 
   >{\centering\arraybackslash}X
   >{\centering\arraybackslash}X
   >{\centering\arraybackslash}X
   >{\centering\arraybackslash}X%replace X with p{2cm} if customizing width
   >{\centering\arraybackslash}X}
  \hline
  \hline
 Device & $t_{TaIrTe_4}$ (nm) & $t_{CGT}$ (nm) & $t_{thBN}$ (nm) & $t_{bhBN}$ (nm) & Electrode geometry \\  
  \hline
  A & 1.6 (Bilayer) & 11.9 & 12.4 & 6.6 & Hall bar\\  
  B & 1.4 (Bilayer) & 8.3 & 20.2 & 5.3 & 12-point disc\\  
  C & 2.4 (Trilayer) & 11.2  & 18.9 & 41.1  & Hall bar\\  
  D & Bilayer & 8.0  & 20.2 & 5.3  & Hall bar\\
  \hline
  \hline
\end{tabularx}
\label{SI Table:deviceinfo}
\end{table}

\begin{figure*}[ht]
\centering
\includegraphics[width=1\textwidth]{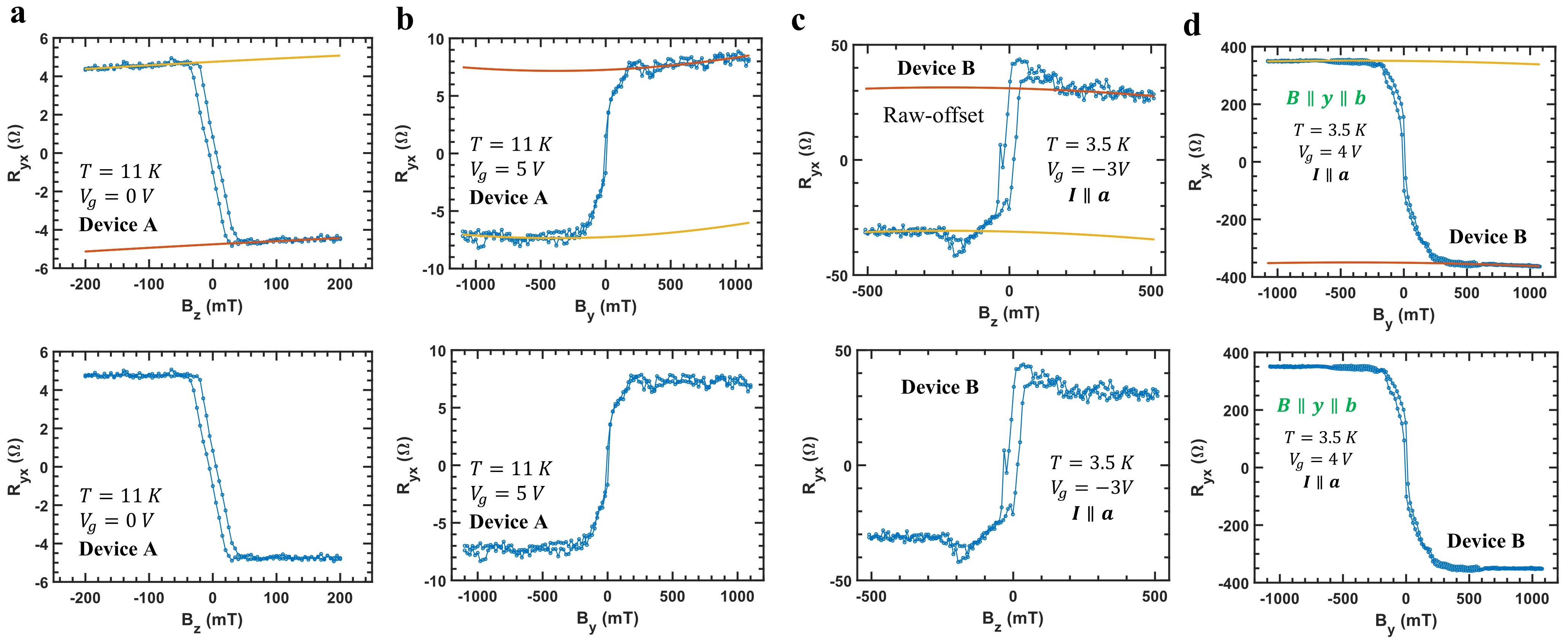}
\caption{\label{SI_Fig:symmetrization}\textbf{Transverse resistance hysteresis loops before magnetic field symmetrization.}
\textbf{a}, Raw transverse resistance ($R_{yx}$) vs $B_z$ in Fig. \ref{Main_Fig2}e before anti-symmetrization relative to the magnetic field with the offset removed (upper panel) and after the removal of non-AHE background up to 2nd order polynomials (lower panel).
\textbf{b}, Raw transverse resistance ($R_{yx}$) vs $B_y$ at $V_g=5~V$ in Fig. \ref{Main_Fig3}d before anti-symmetrization relative to the magnetic field with the offset removed (upper panel) and after the removal of non-AHE background up to 2nd order polynomials (lower panel).
\textbf{c}, Raw transverse resistance ($R_{yx}$) vs $B_z$ in Fig. \ref{Main_Fig4}d before anti-symmetrization relative to the magnetic field with the offset removed (upper panel) and after the removal of non-AHE background up to 2nd order polynomials (lower panel).
\textbf{d}, Raw transverse resistance ($R_{yx}$) vs $B_y$ at $V_g=5~V$ in Fig. \ref{Main_Fig4}e before anti-symmetrization relative to the magnetic field with the offset removed (upper panel) and after the removal of non-AHE background up to 2nd order polynomials (lower panel).}
\end{figure*}

\begin{figure*}[ht]
\centering
\includegraphics[width=1\textwidth]{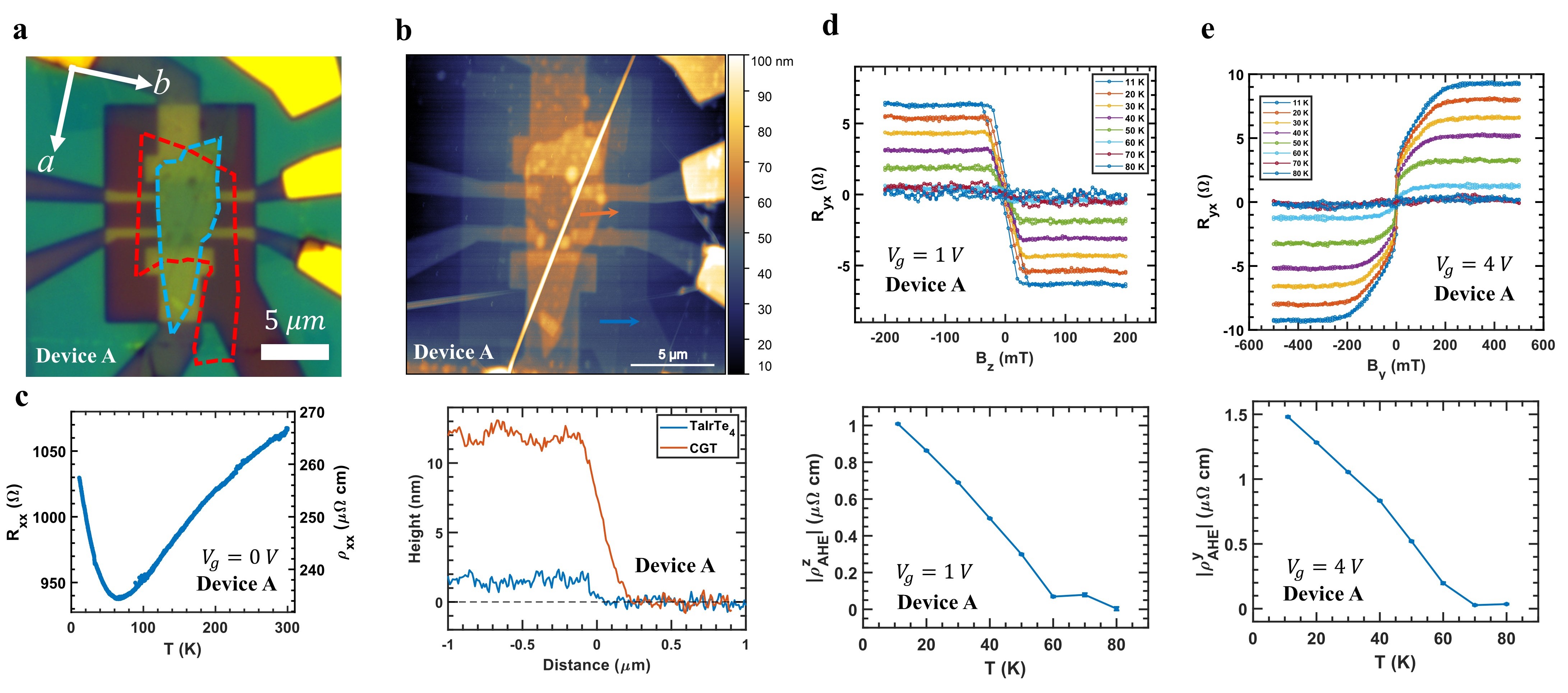}
\caption{\label{SI_Fig:DevA_additional}\textbf{Device A: Characterization, transport property, and additional temperature dependence results}
\textbf{a}, Optical micrograph of device A, where the TaIrTe$_4$ (dashed red) and CGT (dashed cyan) flakes are outlined. In device A, the angle between the \textit{b}-axis and the current channel is $\alpha_b = 78^\circ$.
\textbf{b}, AFM topography map of device A (upper panel) and the height line scans of TaIrTe$_4$ and CGT (lower panel) taken around arrow indications in the map.
\textbf{c}, The longitudinal resistance (resistivity) of device A as a function of temperature.
\textbf{d}, Upper panel: The $R_{yx}$ vs $B_z$ hysteresis loops measured in device A at various temperatures. 
Lower panel: The $m_z$-dependent AHE resistivity magnitude ($|\rho_{AHE}^z|$) in device A as a function of temperature.
\textbf{e}, Upper panel: The $R_{yx}$ vs $B_y$ hysteresis loops measured in device A at various temperatures. 
Lower panel: The $m_y$-dependent AHE resistivity magnitude ($|\rho_{AHE}^y|$) in device A as a function of temperature. Note that in the orientation of device A, $|\rho_{AHE}^y|=\sin\alpha_b |\rho_{AHE}^b|\simeq|\rho_{AHE}^b|$.}
\end{figure*}

\begin{figure*}[ht]
\centering
\includegraphics[width=1\textwidth]{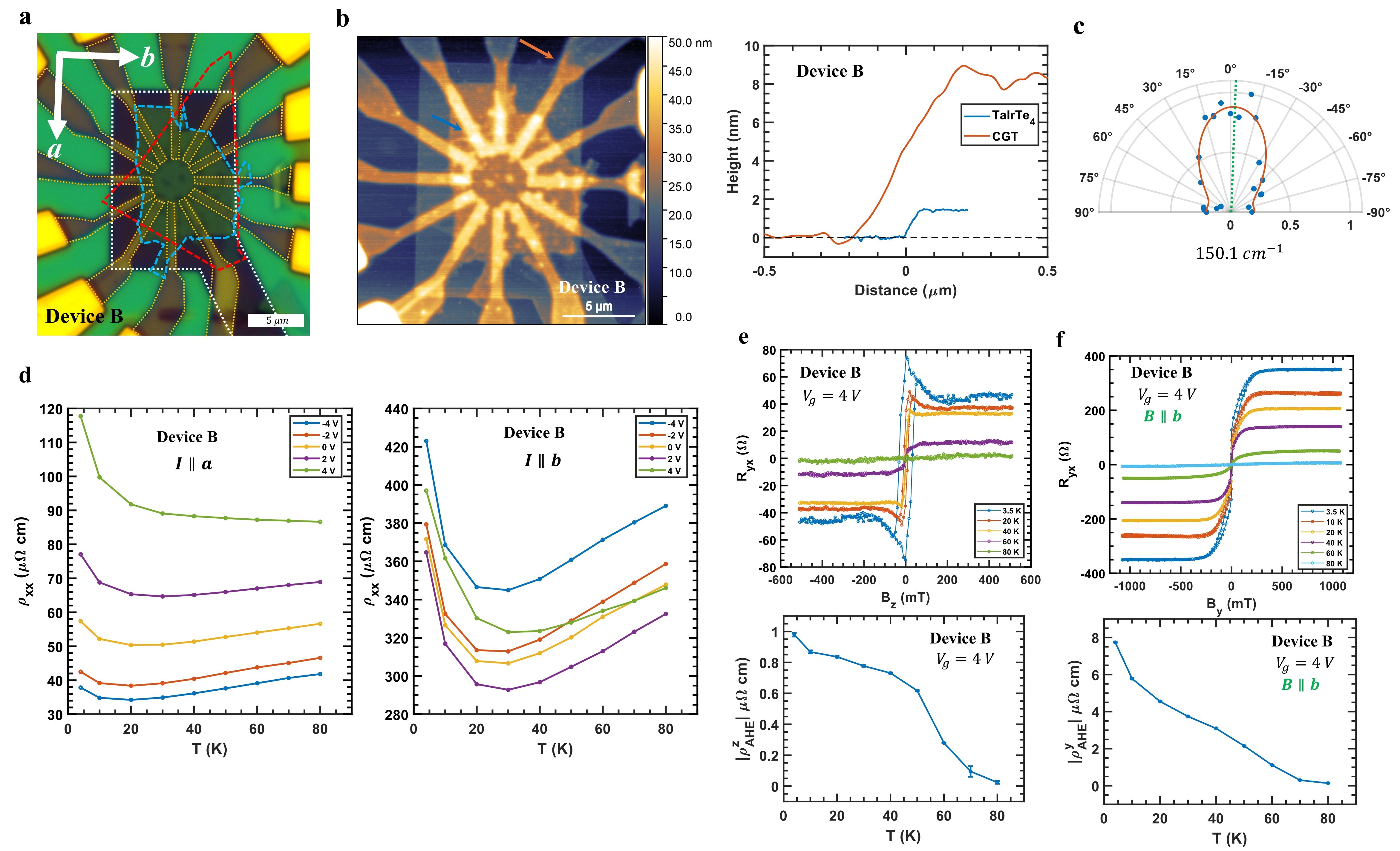}
\caption{\label{SI_Fig:DevB_additional}\textbf{Device B: Characterization, transport property, and additional temperature dependence results}
\textbf{a}, Optical micrograph of device B, where the TaIrTe$_4$ (dashed red), CGT (dashed cyan) flakes, Pt electrodes (dashed yellow), and bottom Pt gate (dashed gray) are outlined. In device B, the angle between the \textit{b}-axis and the current channel is $\alpha_b = 95^\circ$.
\textbf{b}, AFM topography map of device B (left panel) and the height line scans of TaIrTe$_4$ and CGT (right panel) taken around arrow indications in the map.
\textbf{c}, The Angle-dependent polarized Raman spectral intensity at 150.1 cm$^{-1}$ of TaIrTe$_4$ flake relative to the vertical current channel in device B. 
\textbf{d}, The longitudinal resistivity of device B as a function of temperature at various gate voltages when the charge current is applied along the \textit{a}-axis (left panel) and the \textit{b}-axis (right panel) of TairTe$_4$
\textbf{e}, Upper panel: The $R_{yx}$ vs $B_z$ hysteresis loops measured in device B at various temperatures. 
Lower panel: The $m_z$-dependent AHE resistivity magnitude ($|\rho_{AHE}^z|$) in device B as a function of temperature.
\textbf{f}, Upper panel: The $R_{yx}$ vs $B_y$ hysteresis loops measured in device B at various temperatures when $\bm{I\parallel \bm{a}}$ ($\bm{B}\parallel \bm{b}$). 
Lower panel: The $m_y$-dependent AHE resistivity magnitude ($|\rho_{AHE}^y|$) in device B as a function of temperature. Note that in the orientation of device B when $\bm{I\parallel \bm{a}}$, $|\rho_{AHE}^y|=\sin\alpha_b|\rho_{AHE}^b|\simeq|\rho_{AHE}^b|$.}
\end{figure*}

\begin{figure*}[ht]
\centering
\includegraphics[width=1\textwidth]{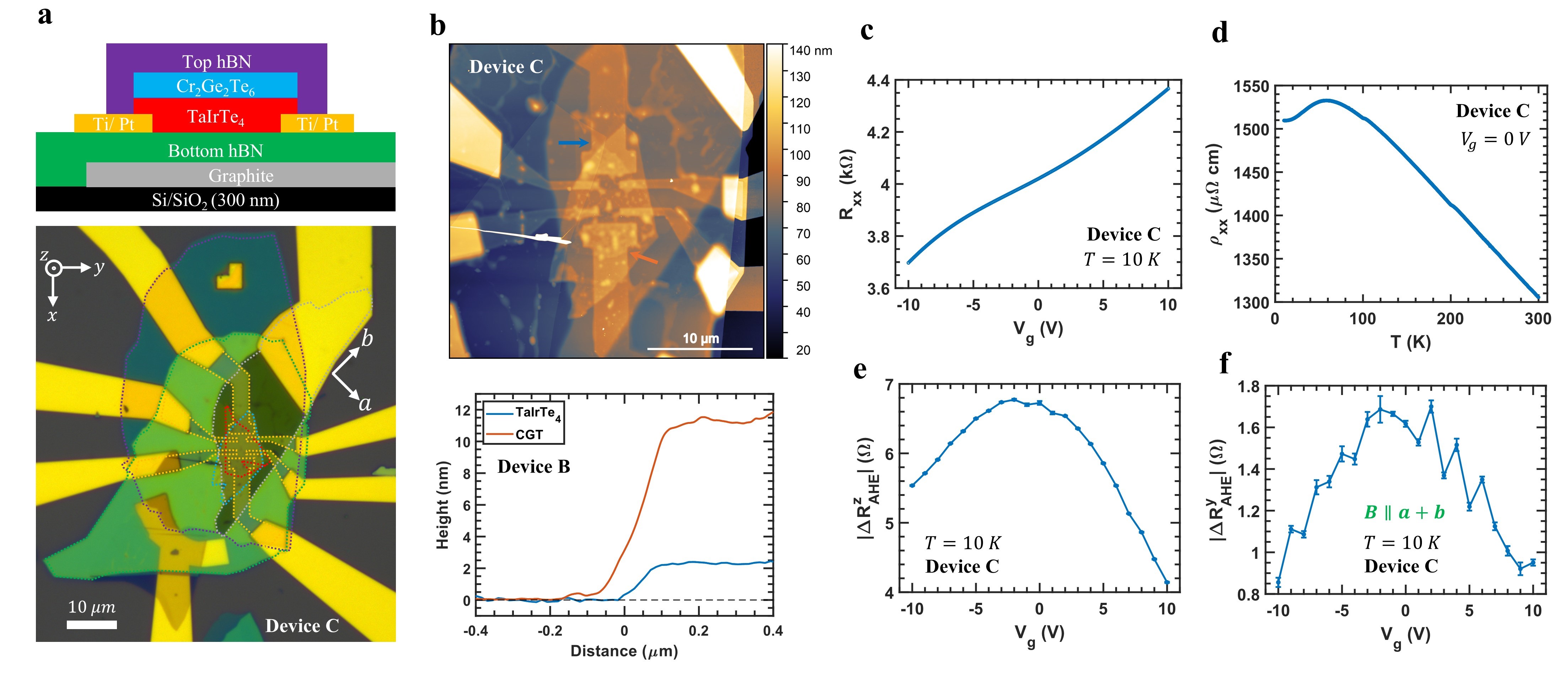}
\caption{\label{SI_Fig:DevC}\textbf{Device C: Characterization, transport property, and gate voltage dependence results}
\textbf{a}, The schematic of the device C side view (upper panel) and the optical micrograph of device C (lower panel), where the TaIrTe$_4$ (dashed red), CGT (dashed cyan), bottom h-BN (dashed green), top h-hBN (dashed purple) flakes, Pt electrodes (dashed yellow), and bottom graphite gate (dashed gray) are outlined. In device C, the angle between the \textit{b}-axis and the current channel is $\alpha_b = 45^\circ$.
\textbf{b}, AFM topography map of device C (upper panel) and the height line scans of TaIrTe$_4$ and CGT (lower panel) taken around arrow indications in the map.
\textbf{c}, The longitudinal resistance of device C as a function of gate voltage at 10 K. 
\textbf{d}, The longitudinal resistivity of device C as a function of temperature.
\textbf{e}, The The $m_z$-dependent AHE resistance magnitude ($|\Delta R_{AHE}^z|$) in device C as a function of gate voltage.
\textbf{f}, The $m_y$-dependent AHE resistance magnitude ($|\Delta R_{AHE}^y|$) in device C as a function of gate voltage. Note that in the orientation of device C, $|\Delta R_{AHE}^y|=\sin\alpha_b|\Delta R_{AHE}^b|\simeq 0.71|\Delta R_{AHE}^b|$.}
\end{figure*}

\begin{figure*}[ht]
\centering
\includegraphics[width=1\textwidth]{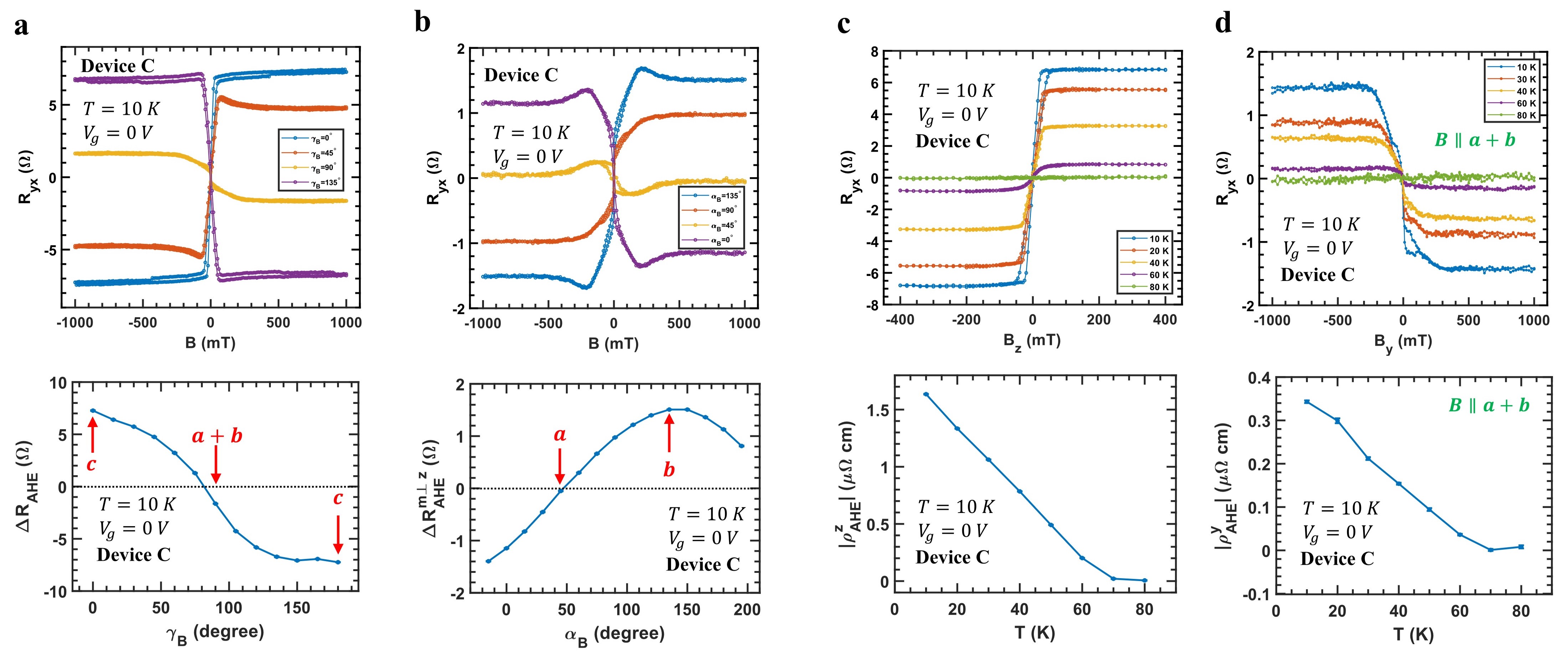}
\caption{\label{SI_Fig:DevC_additional}\textbf{Device C: Magnetic field angle and temperature dependence results}
\textbf{a}, Upper panel: The $R_{yx}$ vs $B$ hysteresis loops measured in device C at various \textit{zy}-plane rotation angle $\gamma_B$.
Lower panel: The total AHE resistance ($\Delta R_{AHE}$) in device C as a function of $\gamma_B$. Note that in the orientation of device C, when $\gamma_B=90^\circ$, $\bm{B}\parallel\bm{a}+\bm{b}$.
\textbf{b}, Upper panel: The $R_{yx}$ vs $B$ hysteresis loops measured in device C at various \textit{xy}-plane rotation angle $\alpha_B$ relative to the current direction (\textit{x}-axis)
Lower panel: The in-plane AHE resistance ($\Delta R_{AHE}^{m\perp z}$) in device C as a function of $\alpha_B$. 
\textbf{c}, Upper panel: The $R_{yx}$ vs $B_z$ hysteresis loops measured in device C at various temperatures.
Lower panel: The $m_z$-dependent AHE resistivity magnitude ($|\rho_{AHE}^z|$) in device C as a function of temperature. 
\textbf{d}, Upper panel: The $R_{yx}$ vs $B_y$ hysteresis loops measured in device C at various temperatures.
Lower panel: The $m_y$-dependent AHE resistivity magnitude ($|\rho_{AHE}^y|$) in device C as a function of temperature. Note that in the orientation of device C, $|\rho_{AHE}^y|=\sin\alpha_b|\rho_{AHE}^b|\simeq 0.71|\rho_{AHE}^b|$.}
\end{figure*}

\begin{figure*}[ht]
\centering
\includegraphics[width=1\textwidth]{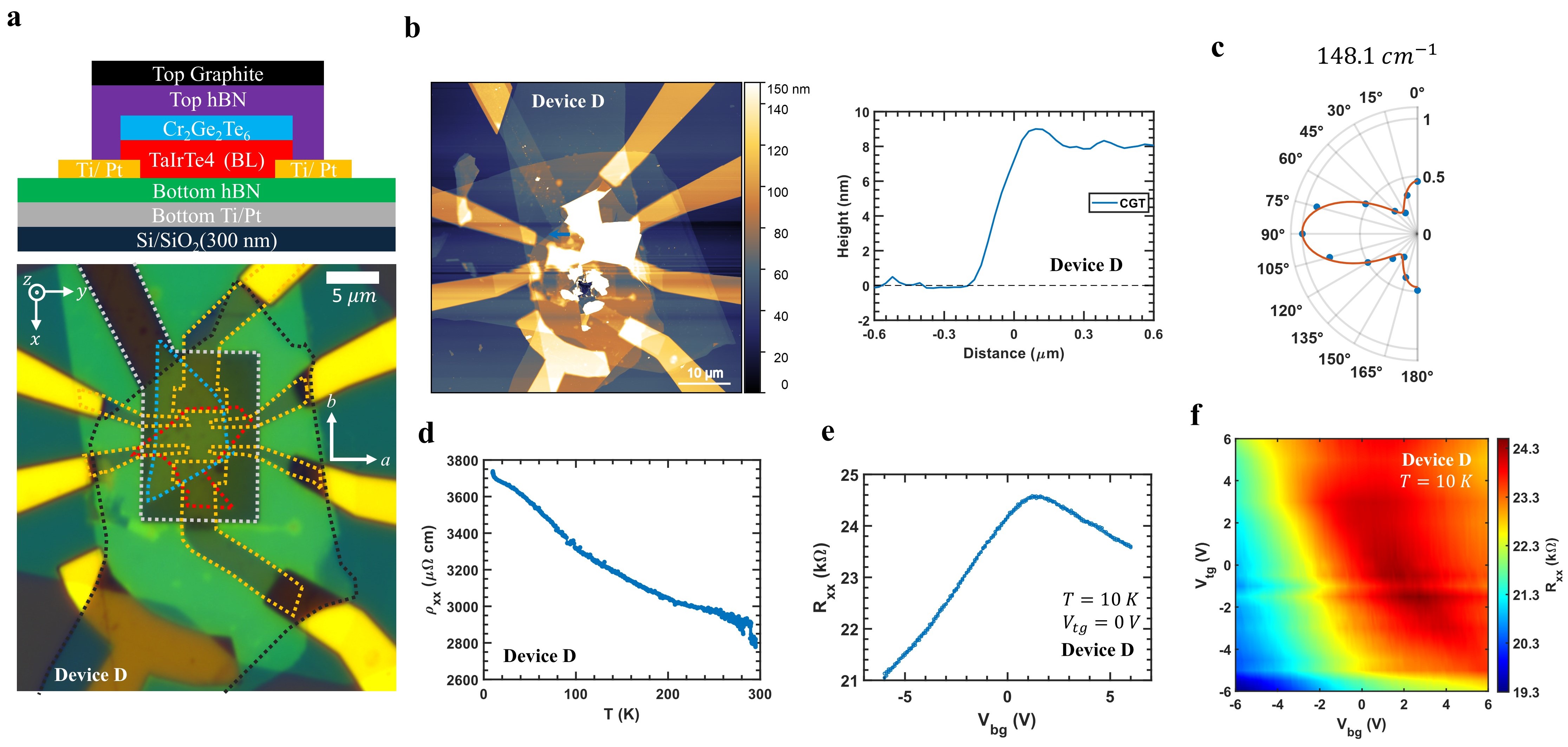}
\caption{\label{SI_Fig:DevD}\textbf{Device D: Characterization and transport property}
\textbf{a}, The schematic of the device D side view (upper panel) and the optical micrograph of device D (lower panel), where the TaIrTe$_4$ (dashed red), CGT (dashed cyan) flakes, Pt electrodes (dashed yellow), bottom Pt gate (dashed gray), and top graphite gate (dashed black) are outlined. In device D, the angle between the \textit{b}-axis and the current channel is $\alpha_b \simeq 0^\circ$.
\textbf{b}, AFM topography map of device D (left panel) and the height line scans of CGT flake (right panel) taken around arrow indications in the map. The device D is damaged after the measurements thus no AFM thickness characterization on TaIrTe$_4$ is performed.
\textit{c}, The Angle-dependent polarized Raman spectral intensity at 148.1 cm$^{-1}$ of TaIrTe$_4$ flake relative to the vertical current channel in device D. 
\textbf{d}, The longitudinal resistivity of device D as a function of temperature. 
\textbf{e}, The longitudinal resistance of device D as a function of the bottom gate voltage ($V_{bg}$) at 10 K when $V_{tg}=0~V$.
\textbf{f}, The longitudinal resistance of device D as a function of the bottom gate voltage ($V_{bg}$) in a range of $V_{tg}$ at 10 K.}
\end{figure*}

\begin{figure*}[ht]
\centering
\includegraphics[width=1\textwidth]{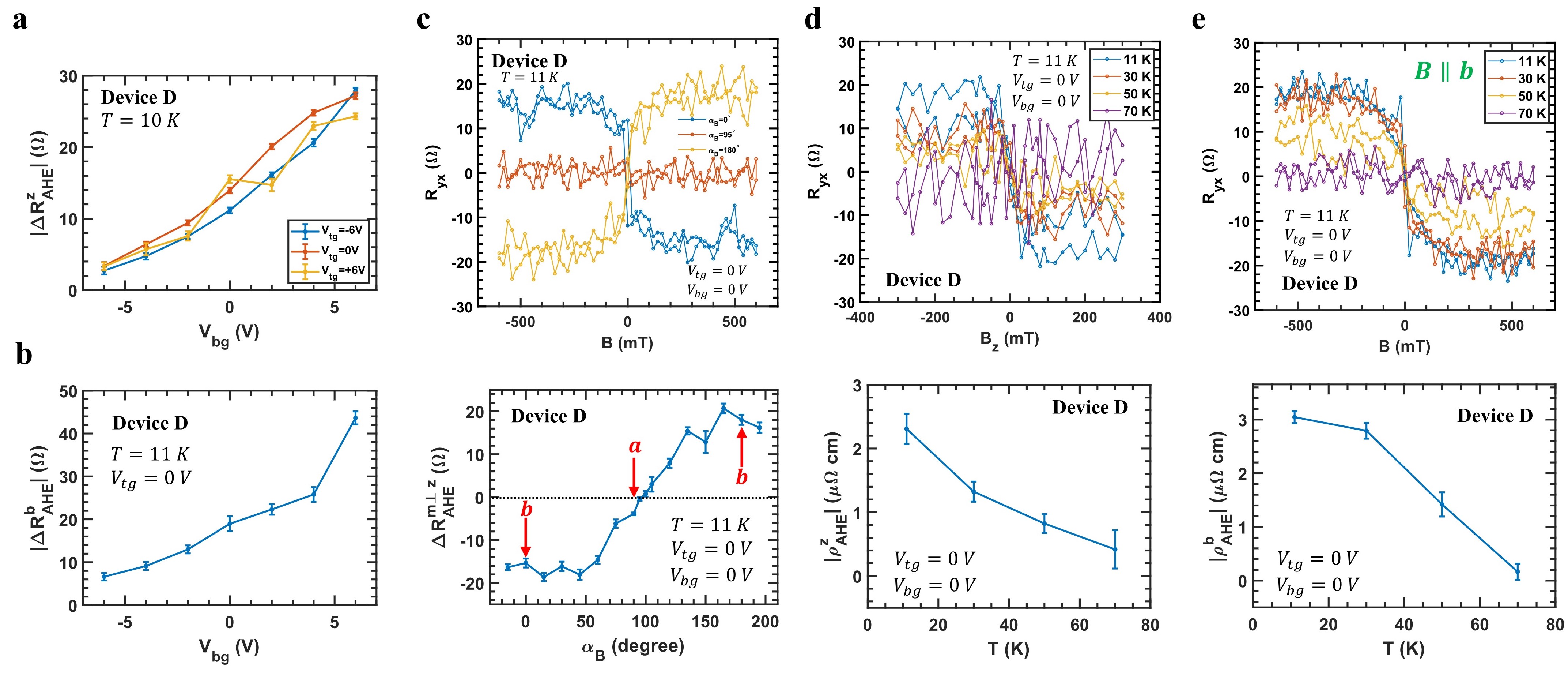}
\caption{\label{SI_Fig:DevD_additional}\textbf{Device D: Gate voltage, magnetic field angle, and temperature dependence results}
\textbf{a}, The $m_z$-dependent AHE resistance magnitude ($|\Delta R_{AHE}^{z}|$) in device D as a function of bottom gate voltage $V_{bg}$ at various top gate voltage $V_{tg}$. 
\textbf{b}, The $m_b$-dependent AHE resistance magnitude ($|\Delta R_{AHE}^{b}|$) in device D as a function of bottom gate voltage $V_{bg}$.
\textbf{c}, Upper panel: The $R_{yx}$ vs $B$ hysteresis loops measured in device D at various \textit{xy}-plane rotation angle $\alpha_B$.
Lower panel: The in-plane AHE resistance ($\Delta R_{AHE}^{m\perp z}$) in device D as a function of $\alpha_B$. 
\textbf{d}, Upper panel: The $R_{yx}$ vs $B_z$ hysteresis loops measured in device D at various temperatures.
Lower panel: The $m_z$-dependent AHE resistivity magnitude ($|\rho_{AHE}^z|$) in device D as a function of temperature. 
\textbf{e}, Upper panel: The $R_{yx}$ vs $B$ ($\bm{B}\parallel \bm{b}$) hysteresis loops measured in device D at various temperatures. 
Lower panel: The $m_b$-dependent AHE resistivity magnitude ($|\rho_{AHE}^b|$) in device D as a function of temperature.}
\end{figure*}

\begin{figure*}[ht]
\centering
\includegraphics[width=1\textwidth]{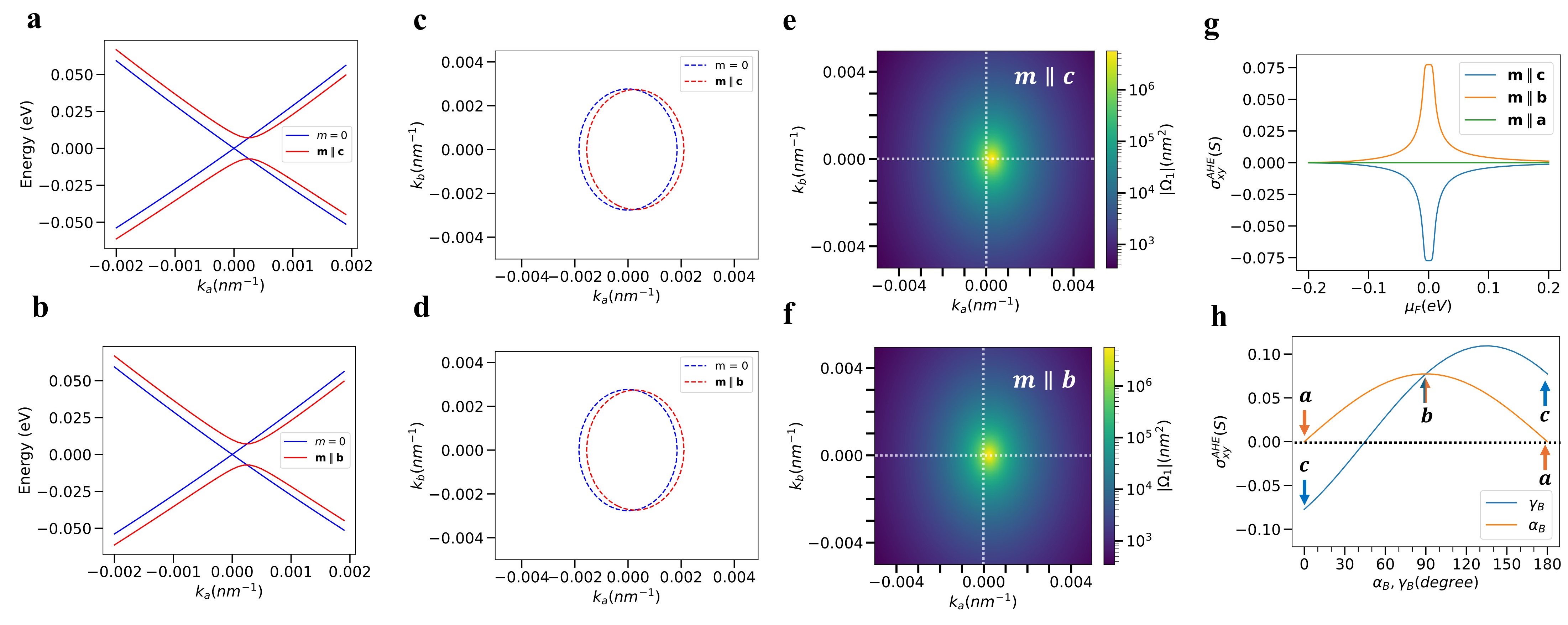}
\caption{\label{SI_Fig:Theory}\textbf{Minimal microscopic model: Unconventional AHE in lower symmetry systems.}
\textbf{a,b,} Band structures for magnetization aligned along the $\bm{c}$-axis (\textbf{a}) and along the $\bm{b}$-axis (\textbf{b}) are shown with the band structure with no magnetization, using the following parameters: $g_{ab} = 20~\mathrm{eV \cdot nm}$, $g_{ba} = -20~\mathrm{eV \cdot nm}$, $g_{ac} = 20~\mathrm{eV \cdot nm}$, $\tilde{g}_{ac} = 10~\mathrm{eV \cdot nm}$, exchange coupling strength $\Delta_{\mathrm{ex}} = 10~\mathrm{meV}$, and effective mass $\tilde{m}_a =\tilde{m}_b= 0.1 m_e$.
\textbf{c,d,} 2D band cut in $k_a$-$k_b$ plane for $\bm{m}\parallel\bm{c}$ (\textbf{c}) and $\bm{m}\parallel\bm{b}$ (\textbf{d}).
\textbf{e,f,} Berry curvature of the valence band for $\bm{m}\parallel\bm{c}$ (\textbf{e}) and $\bm{m}\parallel\bm{b}$ (\textbf{f}). 
\textbf{g}, The calculated anomalous Hall conductivity $\sigma_{xy}^{\mathrm{AHE}}$ as a function of electron chemical potential $\mu_F$ is shown for all three orthogonal magnetization configurations, qualitatively consistent with the experimental observations of AHE in a magnetic system with $\mathcal{C}_s$ symmetry.
\textbf{g}, The calculated anomalous Hall conductivity $\sigma_{xy}^{\mathrm{AHE}}$ is shown as a function of the magnetic field angle when the field $\bm{B}$ is rotated in the \textit{bc}-plane ($\gamma_B$) and \textit{ab}-plane ($\alpha_B$), assuming the magnetization $\bm{m}$ is aligned with the magnetic field ($\bm{m} \parallel \bm{B}$). The angle definitions are as follows: when $\gamma_B = 0^\circ$, $\bm{B} \parallel \bm{c}$; and when $\alpha_B = 0^\circ$, $\bm{B} \parallel \bm{a}$. The calculated results are consistent with the experimental observations of AHE in a magnetic system with $\mathcal{C}_s$ symmetry.
}
\end{figure*}

% \end{appendices}

\end{document}